\documentclass[useAMS,usenatbib]{mn2e}
\usepackage{graphicx}
\usepackage{amsmath}
\usepackage{xcolor}  
\usepackage{float}
\usepackage{color}
\usepackage{amssymb}
\usepackage{gensymb}

\newcommand{\be}{\begin{equation}}
\newcommand{\ee}{\end{equation}}

\newcommand{\ifm}[1]{\relax\ifmmode#1\else$\mathsurround=0pt #1$\fi}
\newcommand{\kms}{\ifmmode\,{\rm km}\,{\rm s}^{-1}\else km$\,$s$^{-1}$\fi}

\newcommand{\ltsima}{$\; \buildrel < \over \sim \;$}
\newcommand{\lsim}{\lower.5ex\hbox{\ltsima}}
\newcommand{\gtsima}{$\; \buildrel > \over \sim \;$}
\newcommand{\gsim}{\lower.5ex\hbox{\gtsima}}

\definecolor{green}{rgb}{0,0.5,0}
\definecolor{grey}{rgb}{0.4,0.5,0.7}

\def\M11{M_{11}}
\def\V100{V_{100}}
\def\R1{R_{Mpc}}
\def\T6{T_6}

\begin{document}

\title[On the  $\Sigma_1$--$M_\star$ quenching boundary]{On the origin of the $\Sigma_1$--$M_\star$ quenching boundary}

\pagerange{\pageref{firstpage}--\pageref{lastpage}} \pubyear{2023}

\author[Cattaneo et al.]{A.~Cattaneo$^{1,2}$, P.~Dimauro$^{3}$, I.~Koutsouridou$^{4}$
\\
\\
$^1$Observatoire de Paris, LERMA, PSL University, 61 avenue de l’Observatoire, 75014 Paris, France \\
$^2$Institut d'Astrophysique de Paris, CNRS, 98bis Boulevard Arago, 75014 Paris, France\\
$^3$INAF—Osservatorio Astronomico di Roma, via di Frascati 33, I-00078 Monte Porzio Catone, Italy\\
$^4$Dipartimento di Fisica e Astronomia, Universit{\`a} degli Studi di Firenze, Via G. Sansone 1, 50019 Sesto Fiorentino, Italy\\
}

\maketitle

\label{firstpage}


\begin{abstract}

We have considered a phenomenologically motivated model in which galaxies are quenched when the energy output of the central black hole exceeds a hundred times the gravitational binding energy of the baryons in the host halo.
The model reproduces the mass functions of star-forming and quiescent galaxies at $0<z<2.5$ and the quenching boundary on a $\Sigma_1$—$M_\star$ diagram. The quenching boundary arises because of the
colour--morphology relation. The stellar surface density $\Sigma_1$ in the central kiloparsec is a morphological indicator. Galaxies becomes redder as  $\Sigma_1$ increases until they cross the quenching boundary and enter the passive population.
Mergers drive the growth of supermassive black holes and the morphological evolution that accompany the migration to the red sequence. That is the origin of the population of high-mass passive galaxies. 
At lower masses, passive galaxies are mainly satellites that ceased to form stars because of environmental effects.

\end{abstract} 


 
\section{Introduction}
\label{Introduction}

The stellar surface density $\Sigma_1$ within the central 1\,kpc of a galaxy increases with the stellar mass $M_\star$: $\Sigma_1\propto M_\star^{0.9}$ for star-forming galaxies and 
$\Sigma_1\propto M_\star^{0.7}$ for passive galaxies  \citep{barro_etal17}. For a given $M_\star$, passive galaxies have a higher $\Sigma_1$ than star-forming galaxies.
The {\it loci} of passive and star-forming galaxies on the $\Sigma_1$—$M_\star$ diagram are separated by a ``quenching boundary” $\Sigma_1\propto M_\star^{0.66}$ \citep{chen_etal20}.

By arguing that $\Sigma_1$ is a proxy for the mass $M_\bullet$ of the central black hole (BH) and that $M_\star$ is a proxy for the virial mass $M_{\rm vir}$ and the virial velocity $v_{\rm vir}$ of the dark matter (DM) halo,
\citet{chen_etal20} proposed that the quenching boundary corresponds to a critical BH mass $M_\bullet^{\rm quench}\propto M_{\rm vir}v_{\rm vir}^2$.
Quenching occurs when the energy deposited by the BH into the surrounding gas ($\propto M_\bullet$) is larger than a fraction or a multiple of the gravitational binding energy of the baryons within the halo ($\propto M_{\rm vir}v_{\rm vir}^2$). The argument has the strength of being entirely empirical. The only theoretical assumption is the cold DM cosmology. Its weakness is that it relies on indirect proxy arguments.

In \citet{koutsouridou_cattaneo22}, we used the GalICS 2.2  semi-analytic model (SAM) of galaxy formation to investigate the consistency of this scenario with other observational data.
We assumed that feedback from the central BH (quasar feedback) blows out all the gas within the inner parts of a galaxy (i.e. the central starburst, the bulge and the bar, but not the disc) as soon as $M_\bullet$ is large enough to satisfy the quenching criterion:
\begin{equation}
\epsilon_{\rm eff}M_\bullet{\rm c}^2>{1\over 2}f_{\rm b}M_{\rm vir}v_{\rm vir}^2,
\label{quenching_crit}
\end{equation}
where c is the speed of light, $f_{\rm b}=0.16$ is the universal baryon fraction and $\epsilon_{\rm eff}= 0.00115$ is a parameter of the SAM calibrated to reproduce the quenching boundary that separates SF and passive galaxies on the $M_\bullet$—$M_\star$ diagram \citep{terrazas_etal16}. 

Massive early-type galaxies have old stellar populations (e.g. \citealp{thomas_etal05}). The physical processes that prevent gas accretion from the circumgalactic medium (CGM) from reactivating star formation today  constitute a different problem (the so-called `maintenance problem') with respect to what caused the initial quenching of star formation at high redshift.
However, in \citet{koutsouridou_cattaneo22} and this article, we assume that, after the initial quenching, gas accretion is permanently shut down. The physics of maintenance  are beyond the scope of our phenomenologically motivated analysis. We can nevertheless give two arguments for the plausibility of our assumption. The first is that quasar feedback heats the CGM to high entropy, so that its cooling time becomes long. The second is that quasar feedback is followed by the onset of a self-regulation mechanism preventing effective cooling (e.g. \citealp{cattaneo_teyssier07}).

With the assumption in Eq.~(\ref{quenching_crit}), \citet{koutsouridou_cattaneo22} could reproduce the evolution of the mass function of galaxies since $z=2.5$ as well as the fraction of passive galaxies at $z=0$ as a function of $M_\star$ for central and satellite galaxies separately.
Our SAM was also in good agreement with the $M_\bullet$—$M_\star$ relation and with the morphological properties of galaxies in the local Universe (i.e. bulge-to-total mass ratios). The current article improves our previous work in three ways. 

1) In  \citet{koutsouridou_cattaneo22}, the comparison with the observations was limited to $z=0$ except for the evolution of the total mass function of galaxies.
Here, we consider the evolution of the mass functions of star-forming and passive galaxies separately and we extend the comparison to other observables
such as the radii of discs and bulges. Radii are important for the calculation of $\Sigma_1$ (see below).

2) \citet{chen_etal20} started from the $\Sigma_1$—$M_\star$ relation and inferred a  quenching criterion of the form in Eq.~(\ref{quenching_crit}).
Here we close the circle  by demonstrating that a model based on the quenching criterion in Eq.~(\ref{quenching_crit})
reproduces \citet{chen_etal20}'s quenching boundary on the $\Sigma_1$—$M_\star$ diagram. The test is not trivial because Eq.~(\ref{quenching_crit})
contains no condition on $\Sigma_1$.

3) For \citet{chen_etal20}, the correlation between $M_\bullet$ and $\Sigma_1$ was purely empirical (see also \citealp{sahu_etal22} for direct observational evidence from galaxies with dynamical BH mass estimates). 
With GalICS~2.3,  we can explore its  astrophysical origin.

The structure of the article is as follows. In Section~2, we introduce our SAM and the novelties of the  GalICS~2.3 version used for this article.
In Sections~3 and 4, we present our results for the stellar mass function of galaxies and $\Sigma_1$, respectively.  Section~5 summarises our conclusions.
In Appendix~A, we discuss the impact that the mass resolution of the N-body simulation used to construct the DM merger trees could have on our results.

\section{The model}
\label{model} 

A systematic presentation of our SAM would take several pages and detract from the focus of the article.
Here we point to the relevant literature and focus on the differences with previous versions.

\citet{cattaneo_etal20} gave a systematic presentation of the GalICS~2.1 version (cosmology, merger trees, DM substructures, merger rates,
accretion of gas onto galaxies and haloes, supernova feedback, galactic fountain) while referring to \citet{cattaneo_etal17} for the internal structure of galaxies (density distributions, radii and characteristic speeds of discs and bulges), star formation rates (SFRs), and morphological transformations, where there had not been any changes since GalICS~2.0 (i.e. \citealp{cattaneo_etal17}). \citet{koutsouridou_cattaneo19} improved GalICS 2.0 
by introducing a model for ram-pressure and tidal stripping.

\subsection{Quenching}

The main novelty of GalICS~2.2 \citep{koutsouridou_cattaneo22} with respect to the previous versions 2.0 and  2.1 was the introduction of a {\it quenching} mechanism to suppress star formation in massive galaxies.
In \citet{koutsouridou_cattaneo22}, this mechanism could take two forms: halo quenching, where gas accretion onto galaxies is shut down above a critical halo mass (model~A), and BH quenching based on Eq.~\ref{quenching_crit} (model~B).

Model~A and~B use the same shock stability criterion, described in \citet{cattaneo_etal20} and further discussed in \citet{tollet_etal22}, to discriminate whether gas is accreted in the cold mode or the hot mode.
The only difference with respect to accretion is the prescription used to decide at which point gas accretion onto galaxies is shut down. Once that happens, in both models: all the gas in the cold CGM is heated to the virial temperature; gas that accretes onto the halo is automatically incorporated into the hot CGM (the intracluster medium in the case of groups and clusters); and the hot CGM is no longer allowed to cool.

Model~A and~B also use the same prescriptions to grow supermassive BHs. The only difference is that model~A contains no BH feedback. In model~A, the growth of BHs is purely limited by how much gas is supplied to them.
The quenching model assumed in the current GalICS~2.3 version is the same as in model~B of \citet{koutsouridou_cattaneo22}.

\subsection{Morphological transformations in mergers}

 The other novelty of GalICS~2.2 was the presence of two options for morphological transformations in galaxy mergers. We referred to them as models~1 and 2.

Model~1  was the same as in previous versions. It assume  a very sharp distinction between major and minor mergers (mergers with mass ratios $\mu$ greater and smaller than 1:4, respectively). Only major mergers transform discs into bulges, trigger starbursts, and feed the growth of supermassive BHs. 

Model~2 is based on hydrodynamic simulations by \citet{kannan_etal15} and assumes a more continuous transition. Mergers transfer a mass fraction $\mu$ of the disc stars to the bulge and a mass fraction $(1-f_{\rm gas,\,disc})\mu$ of the gas in the disc to the central starburst ($f_{\rm gas,\,disc}$ is the gas fraction in the disc). Hence, major mergers do not destroy discs entirely (especially gaseous discs), and minor mergers, too, can contribute the growth of bulges and BHs.
The only difference between major and minor mergers is that major mergers are assumed to trigger starbursts throughout the discs and not only in the central regions of galaxies (\citealp{koutsouridou_cattaneo22} for details).
Model~2 is the default option in GalICS~2.3.

Other minor differences between GalICS~2.2 and the previous versions 2.0 and 2.1 were a new model for disc instabilities (based on numerical simulations by \citealp{devergne_etal20}), a recalibration of the  parameters for supernova feedback, and the star formation timescale in discs.
The GalICS~2.3 version used for this article is almost identical to the B2 variant of GalICS~2.2, which uses model~B for quenching and model~2 for mergers. The only differences are in the density profiles and sizes of bulges (Section~2.3) and in the star formation law for discs (Section~2.4). We have also improved our model for tidal stripping by discovering and correcting a mistake by a factor of two (Section~2.5).

\begin{figure*}
\begin{center}
\includegraphics[width=0.32\hsize]{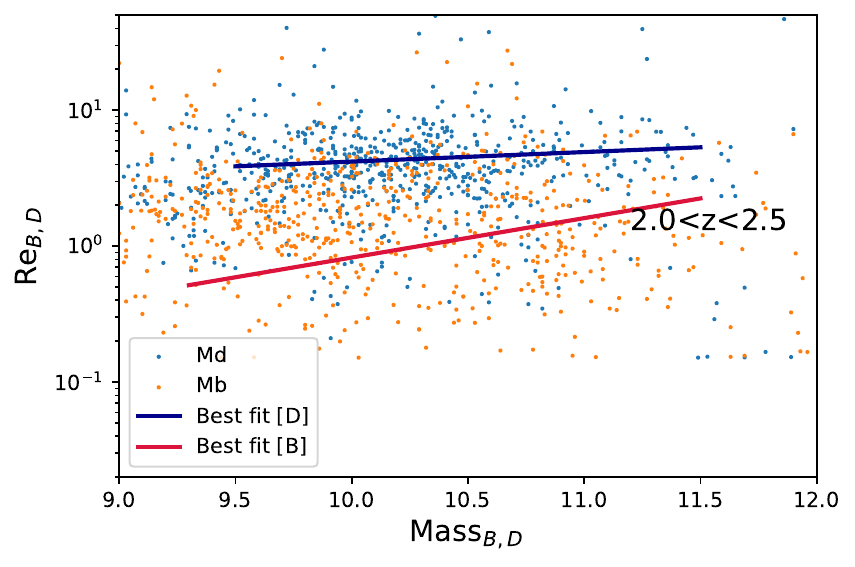} 
\includegraphics[width=0.32\hsize]{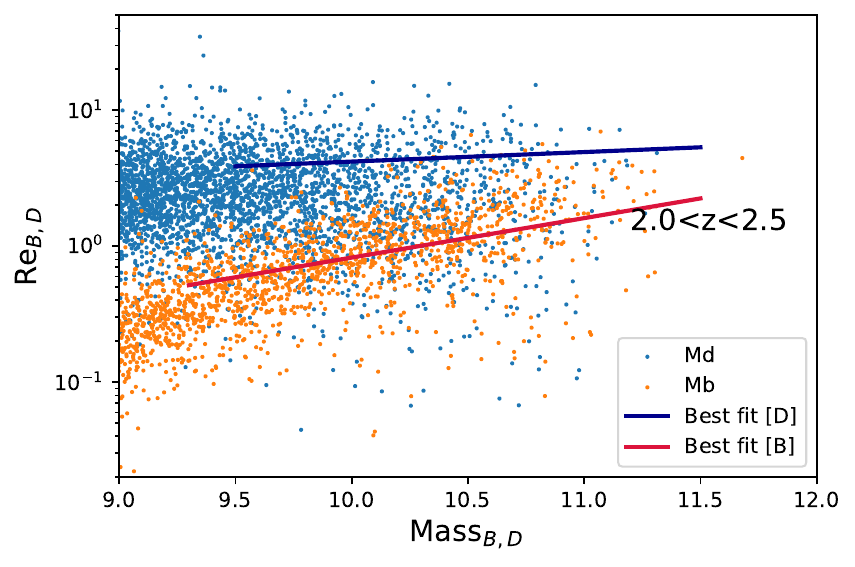} 
\includegraphics[width=0.32\hsize]{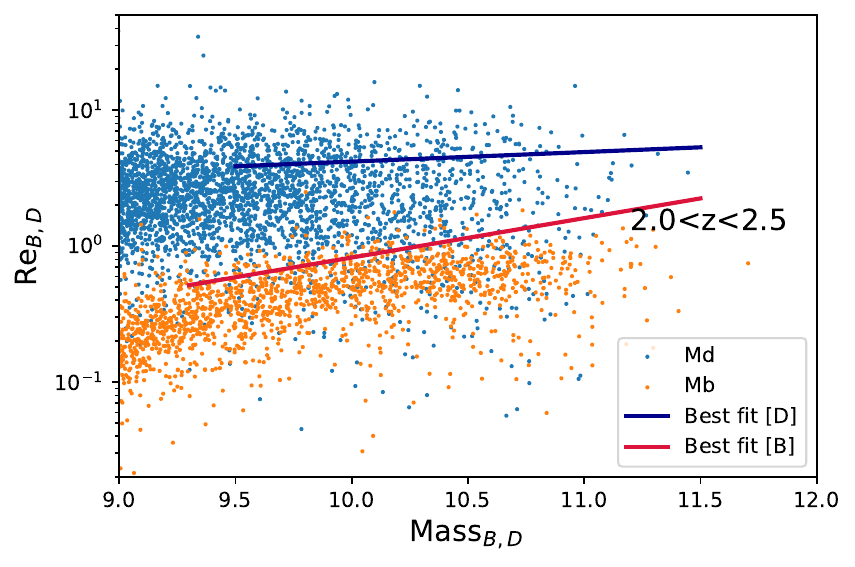}\\
\includegraphics[width=0.32\hsize]{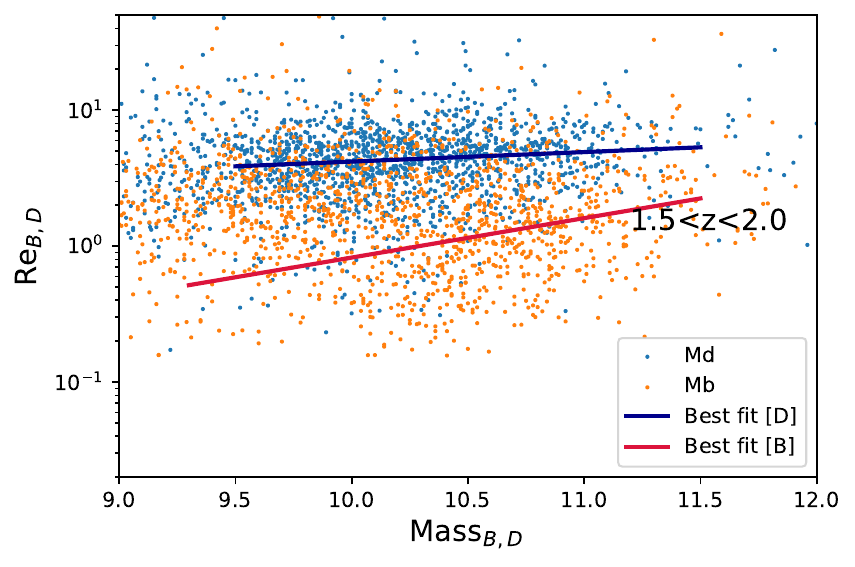} 
\includegraphics[width=0.32\hsize]{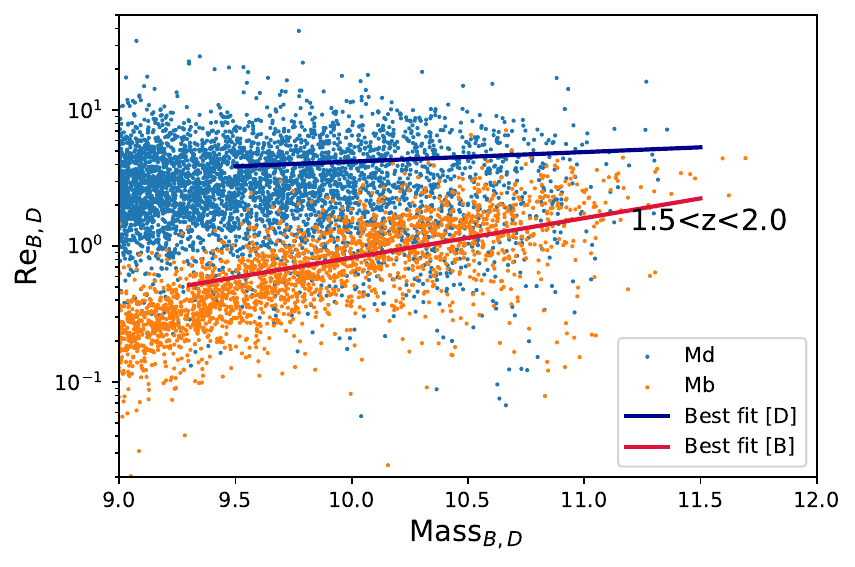} 
\includegraphics[width=0.32\hsize]{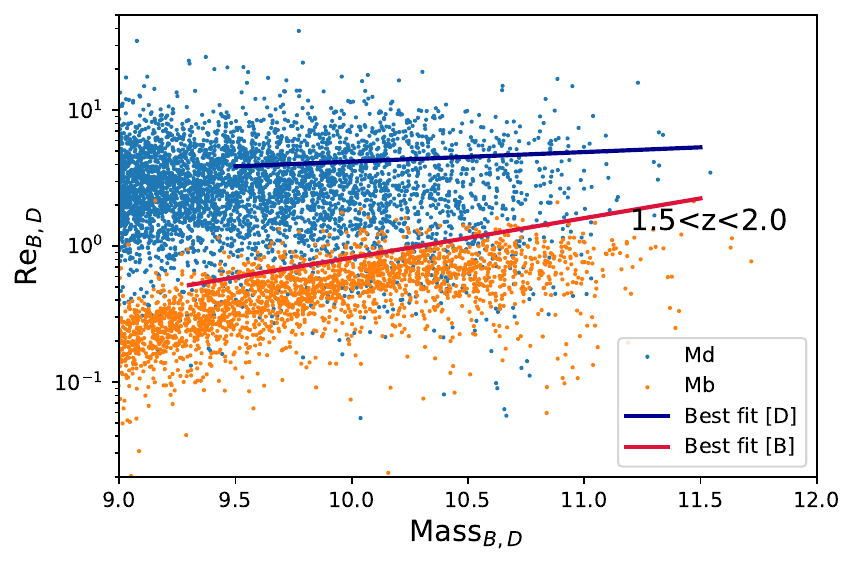}\\
\includegraphics[width=0.32\hsize]{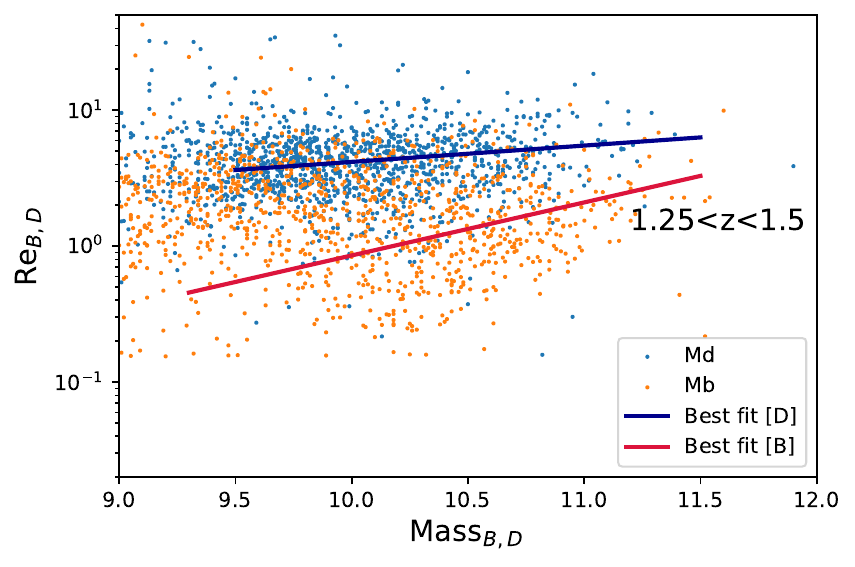} 
\includegraphics[width=0.32\hsize]{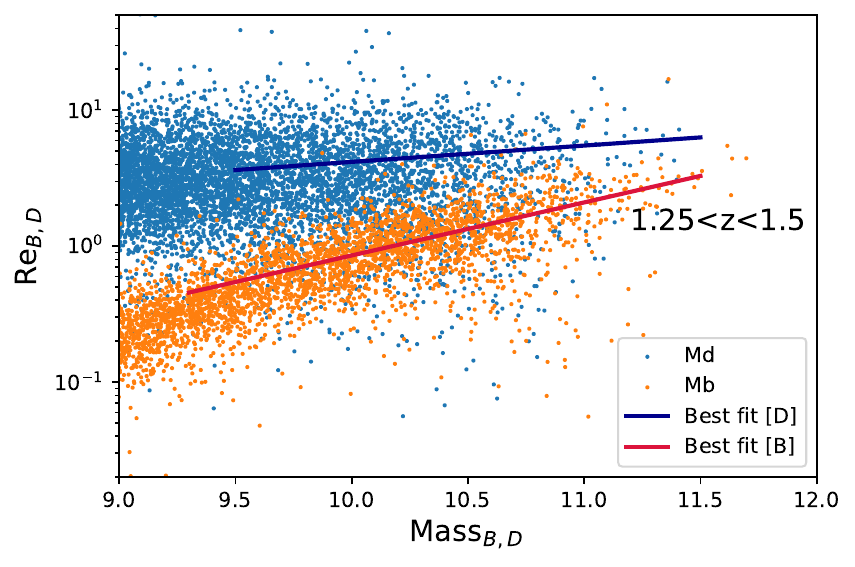} 
\includegraphics[width=0.32\hsize]{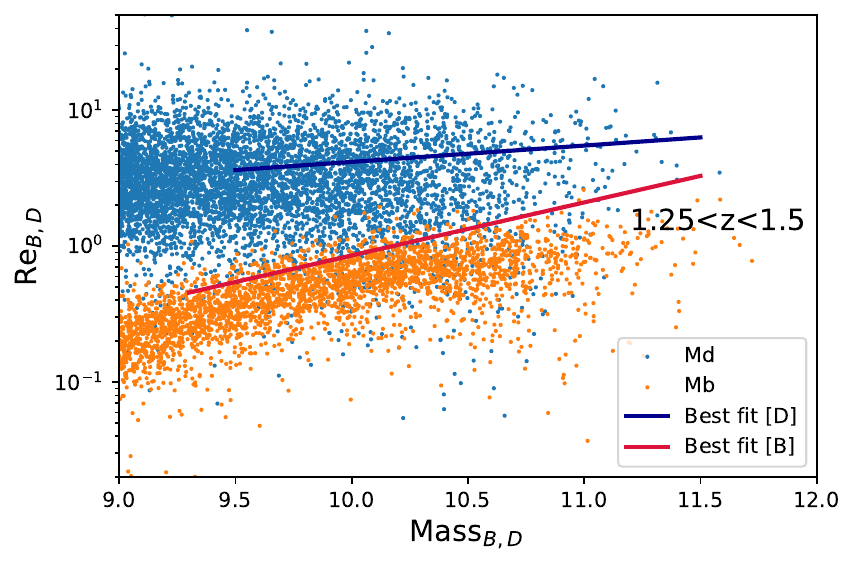}\\ 
\includegraphics[width=0.32\hsize]{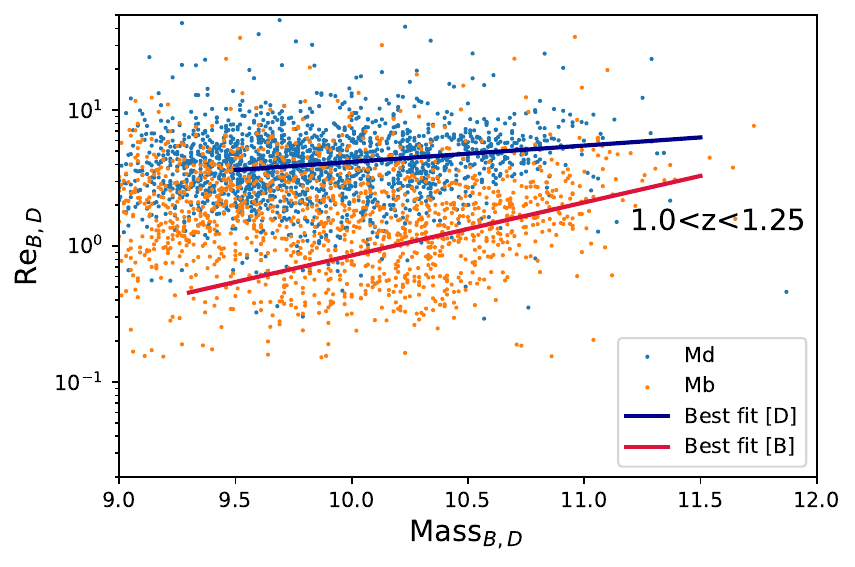} 
\includegraphics[width=0.32\hsize]{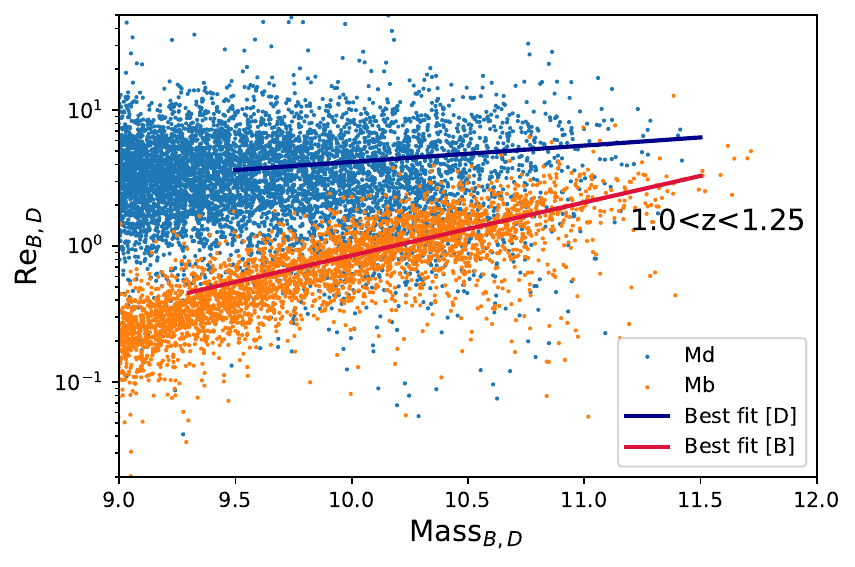} 
\includegraphics[width=0.32\hsize]{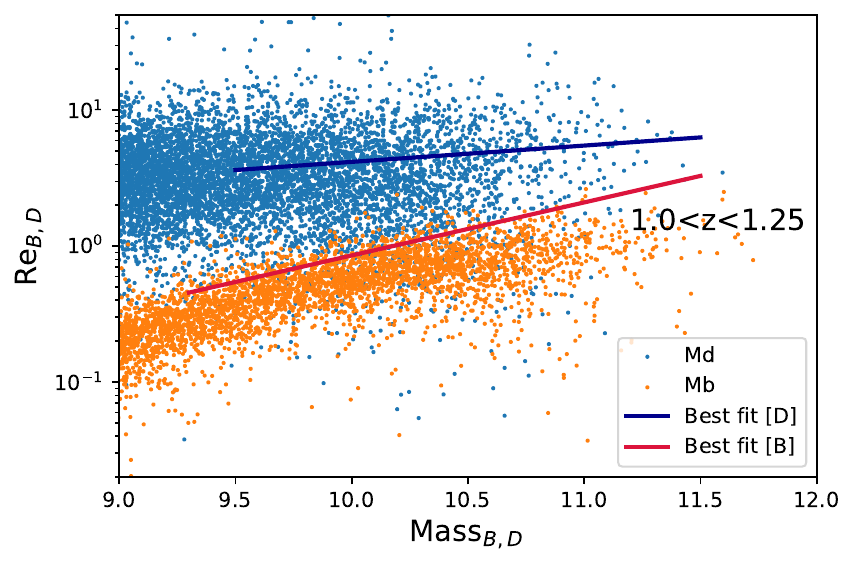}\\ 
\includegraphics[width=0.32\hsize]{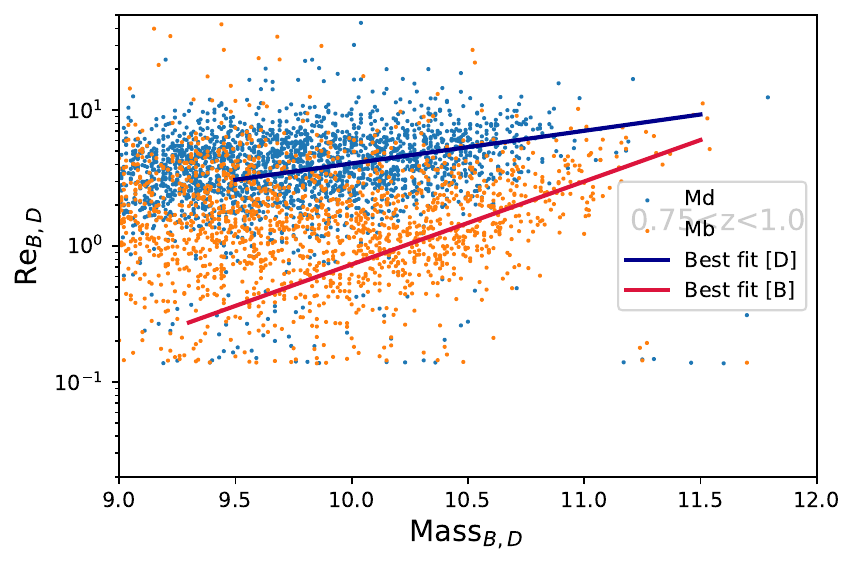} 
\includegraphics[width=0.32\hsize]{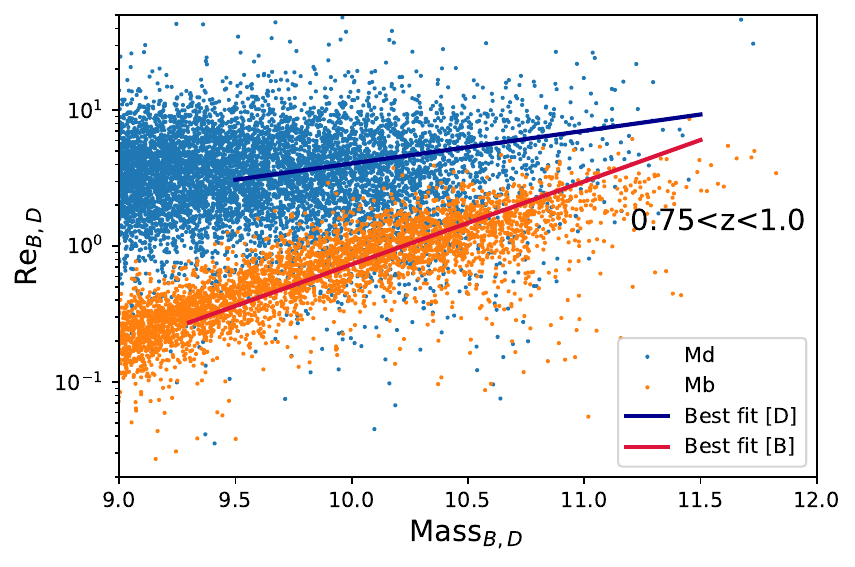} 
\includegraphics[width=0.32\hsize]{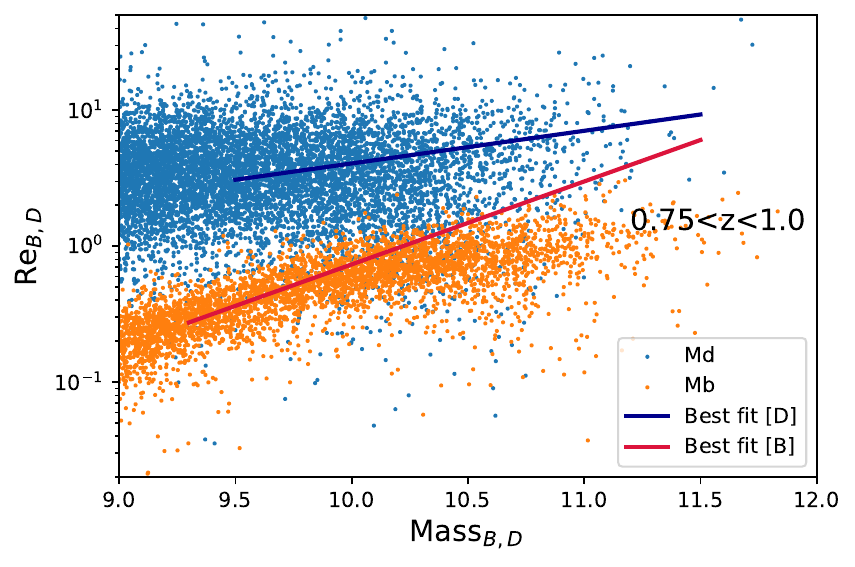}\\ 
\includegraphics[width=0.32\hsize]{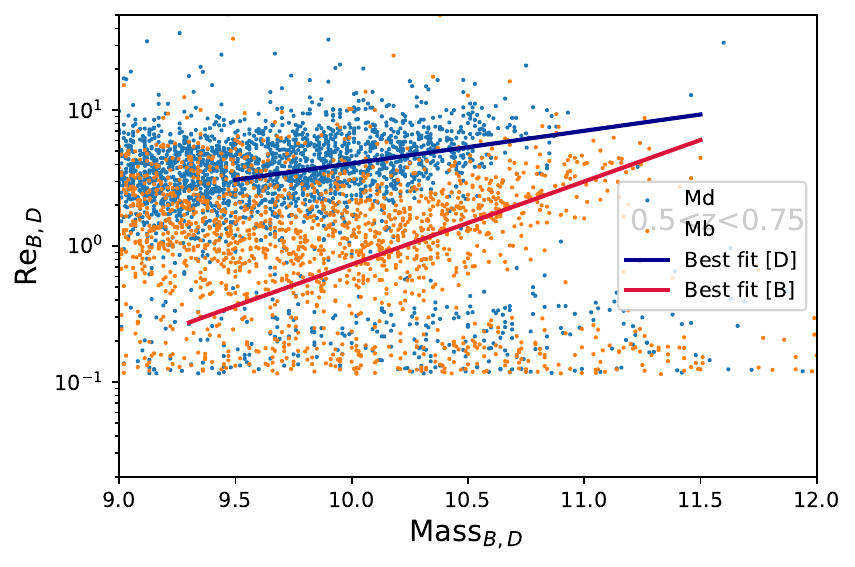}
\includegraphics[width=0.32\hsize]{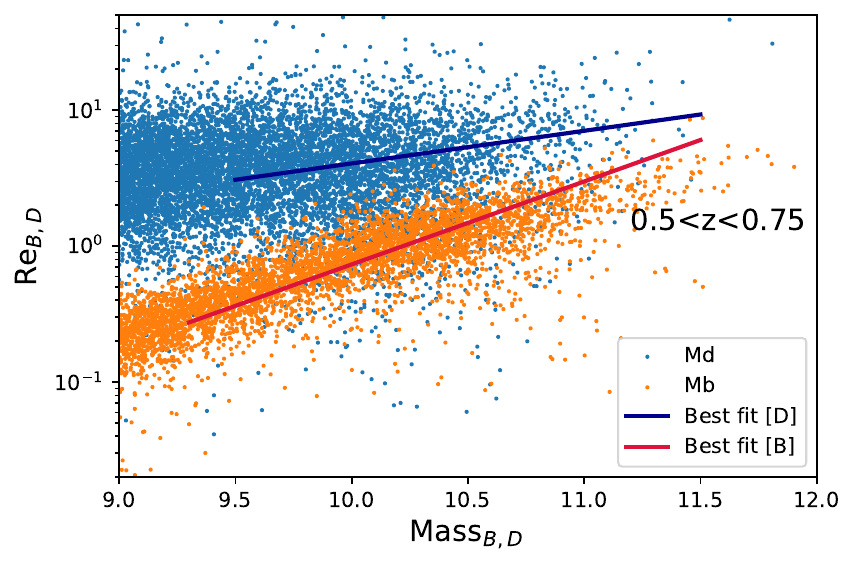}
\includegraphics[width=0.32\hsize]{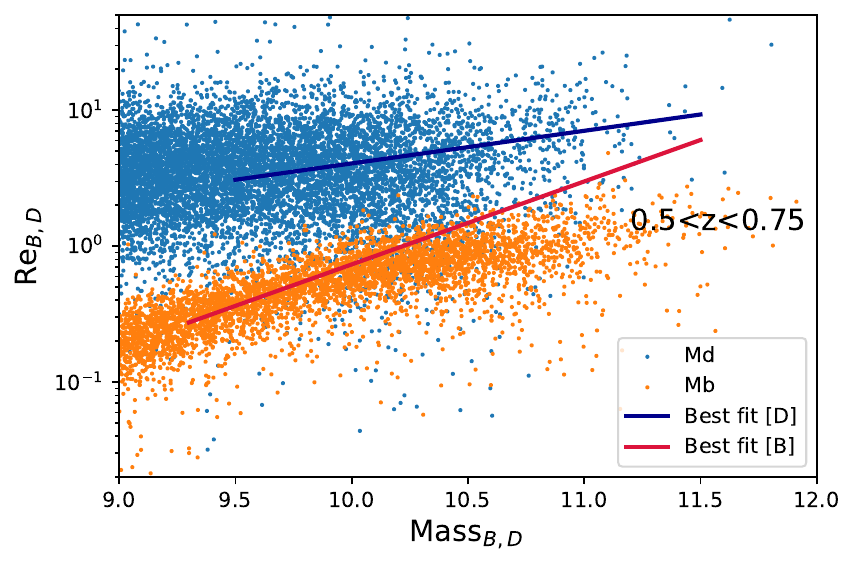}
\end{center}
\caption{Stellar mass--size relation for discs (blue points) and bulges (orange points) in CANDELS ({\it left}), GalICS~2.3 without dissipation ({\it centre}), and GalICS~2.3 with $f_0=0.3$ ({\it right}).
The effective radii $R_{\rm e}$ are half-mass radii on a 2D projection (face-on for discs). The blue and red lines show the mean relations in CANDELS  \citep{dimauro_etal19} and are the same on all three columns at a given $z$.}\label{sizes}
\end{figure*}

\subsection{Density profile and sizes of bulges}

Computing $\Sigma_1$ requires a model for the  distribution of stars in galaxies, especially for the bulge component, which dominates the central region.
Observationally, bulges are fitted with a \citet{Sersic1963} three-parameter profile. Many bulges exhibit central light excesses (cusps) or deficits (cores) with respect to a {S{\'e}rsic} fit  \citep{faber_etal97,Kormendy2009}. The central light excesses of cuspy ellipticals
are assumed to originate from gas that falls into the central regions of galaxies during dissipative mergers and triggers central starbursts (e.g. \citealp{faber_etal97,hopkins_etal09}).

In our SAM, galaxies comprise four components: the disc, the bar, the bulge, and the cusp, which is present only if the last merger was dissipative.
The cusp contains the stars formed in the central starburst at the last major mergers. The bulge corresponds to the pre-existing stellar population. 

One of the main difficulties concerning the structural properties of galaxies is to determine the density distributions of the bulge and the cusp.
We have explored a number of different models including one with two {S{\'e}rsic} profiles, one for the bulge and one for the central cusp
(in addition to the exponential profile used to model the disc). However, introducing more free parameters than constraints just brings further uncertainty into the model. In the end, 
the model that works best at reproducing the sizes of bulges is the simplest one. 
In the current GalICS~2.3 version, we model the bulge and the cusp with a single \citet{hernquist90} profile, so that the cusp is simply the innermost part of the bulge.
The only purpose of maintaining a separate cusp component is to keep track of the stars formed during the last merger.

The Hernquist profile has two parameters.
Hence,   the mass $M_{\rm b}$ and the energy $E_{\rm b}$ of a bulge entirely determine
its density distribution.
The scale-length $a$ of the Hernquist profile is related to $M_{\rm b}$ and $E_{\rm b}$ by the equation  \citep{hernquist90}:
\begin{equation}
E_{\rm b}=-{{\rm G}M_{\rm b}^2\over 12a}.
\label{hernquist_energy}
\end{equation}
Eq.~(\ref{hernquist_energy}) follows from the virial theorem (the total energy is half the gravitational potential energy) and the assumption that there is not much DM or disc matter inside the bulge,
so that one can consider the bulge self-gravitating and  compute its gravitational potential energy from its density distribution alone.

The main interest of the Hernquist profile, besides its simple analytic form, is that its two-dimensional projection closely approximates a \citet{devaucouleurs48} profile with effective radius $R_e\simeq 1.8153a$ (the de Vaucouleurs profile is a  {S{\'e}rsic} profile with index $n=4$). Once we find $E_{\rm b}$, it is straightforward to compute 
\begin{equation}
R_e=-0.151{{\rm G}M_{\rm b}^2\over E_{\rm b}}.
\label{Re}
\end{equation}
and thus the surface density profile of the bulge.

We compute the mechanical energy $E_{\rm b}$ of a bulge formed after a merger event by 
assuming that the baryons that end up in the bulge conserve their mechanical energy:
\begin{equation}
E_{\rm b}=\epsilon_{\rm b1}E_{\rm b1}+\epsilon_{\rm d1}E_{\rm d1}+ \epsilon_{\rm b2}E_{\rm b2}+\epsilon_{\rm d2}E_{\rm d2}.
\label{bulge_energy}
\end{equation}
Here $E_{\rm b1}$, $E_{\rm b2}$, $E_{\rm d1}$, $E_{\rm d2}$  are the mechanical energies of the bulges and the discs of the merging galaxies.
The subscripts 1 and 2 refer to  the larger and the smaller galaxy, respectively.
The coefficients  $\epsilon_{\rm b1}$, $\epsilon_{\rm b2}$, $\epsilon_{\rm d1}$ and $\epsilon_{\rm d2}$ correspond to the fractions of the baryons in the bulge of galaxy 1, the bulge of galaxy 2, the disc of galaxy 1 and the disc of galaxy 2 that end up contributing to the stellar mass of the final bulge, respectively.
We assume that all the stars in the bulge of larger galaxy remain in the bulge component ($\epsilon_{\rm b1}=1$).
 The other three coefficients are computed following the prescriptions in \citet{koutsouridou_cattaneo22}.
Eq.~(\ref{bulge_energy}) does not contain any interaction term because we have assumed that the merging galaxies fall onto each other while starting from rest at infinity  (as in the previous versions of our SAM).

The mechanical energies $E_{\rm b1}$ and $E_{\rm b2}$ of the bulges of the merging galaxies are computed from their masses and sizes with Eq.~(\ref{hernquist_energy}).
The mechanical energies $E_{\rm d1}$ and $E_{\rm d2}$ of their discs are computed from the virial theorem by assuming that the mechanical energy $E_{\rm d}$ of a disc  is the opposite of its kinetic energy:
\begin{equation}
E_{\rm d}=-2\pi\int_0^\infty {1\over 2}v_{\rm c}^2(r)\Sigma_{\rm d}(r) r{\rm\,d}r,
\label{disc_energy}
\end{equation}
where $\Sigma_{\rm d}(r)$ is the disc's surface density profile (a decreasing exponential) and $v_{\rm c}(r)$ is the circular velocity (computed as in \citealp{cattaneo_etal17}).

Rigorously, even in dissipationless (``dry") mergers, the conservation of energy and the virial theorem apply to the whole system and not to its individual components (e.g. the stars or the DM). 
To understand the impact that this may have on our results, we have looked at the {GalMer} database of merger simulations \citep{chilingarian_etal10}.
In dissipationless simulations, the stars transfer to the DM $2$ per cent of their energy in the worst case scenario.
The accuracy of applying the virial theorem to the stars alone is measured by $2T_*/|U_*|$, where $T_*$ is the kinetic energy of the stars and $U_*$ is their gravitational potential energy,
which is due not only to the stars' self-gravity but also to the DM's gravitational potential. Eq.~(\ref{disc_energy}) assumes $2T_*/|U_*|=1$.  In GalMer, $0.80<2T_*/|U_*|<0.95$.
Our approximations do introduce errors at the 20 per cent level but such errors are immaterial when considering the large scatter in both observed and modelled sizes (Fig.~\ref{sizes}).

In dissipative (``wet") mergers, there is the additional complication that the gas that ends up in the cusp radiates before it forms stars.
Hence, Eq.~(\ref{bulge_energy}) overestimates $E_{\rm b}$ and the scale-length $a$ of the Hernquist profile.
\citet{hopkins_etal09a} run hydrodynamic simulations of the formation of bulges in major mergers. They found that the presence of gas reduced the bulge size by a factor of $(1+f_{\rm c}/f_0)^{-1}$,
where $f_{\rm c}$ is the fraction of stellar mass formed during the mergers (i.e. the ratio of the cusp mass to the total bulge mass) and $f_0\sim 0.3$.

Fig.~\ref{sizes} compares the mass--size relations for discs (blue points) and bulges (orange in points) in the CANDELS data (left), in GalICS~2.3 without dissipation (centre) and in GalICS~2.3  with $f_0=0.3$ (right).
The blue and red lines are fits to the observational data for discs and bulges, respectively. They have been reproduced on the central column and the right column to assist the comparison with the observations, but one should note the important scatter of the data points around the mean relations.

In both the dissipationless and the dissipational model, the evolution of $R_{\rm e}(M_\star)$ with $z$ is weaker in GalICS~2.3 than in CANDELS.
The dissipationless model (centre) is in reasonably good agreement with the data at all redshifts. Only the most massive galaxies have bulges that are too large at $z>1.5$ and too small at $z<1$.
Adding dissipation  (Fig.~\ref{sizes}, right) reduces the sizes of the bulges formed in high-$z$ gas-rich mergers and brings GalICS~2.3 in better agreement with the CANDELS data at $z>1.5$.
Mergers at low $z$ have lower gas fractions.  Hence, one would expect that their sizes should be less affected but that is not true because smaller progenitors result in smaller descendants.
Dissipation improves the agreement with the data at $z>1.5$ but spoils it at lower $z$ (right column).
The dissipationless model is the one that reproduces the sizes of galaxies better
across the entire redshift range $0<z<2.5$ and will thus be our reference model in the rest of the article.

There are two possible explanations why the crude model without dissipation works better.
First, in real mergers a lot of the gas is in dense molecular clouds that hydrodynamic simulations cannot resolve and that behave like dissipationless N-body particles.
Second, by assuming a Hernquist profile we have implicitly assumed that all bulges have the same {S{\'e}rsic} index $n=4$, while we know that is not true and that there are 
systematic trends of $n$ with $M_\star$ \citep{kormendy_etal09}.

\subsection{Star formation timescales}

Our SAM distinguishes between quiescent star formation in discs and bursty star formation in cusps and bars.
This distinction admits one exception. In remnants of major mergers, all the gas is starbursting, including that in the disc component\footnote{In GalICS~2.3, major mergers do not destroy discs entirely (Section~2.2).}. 
Merger-driven starbursts continue until they have depleted all the gas present in the merging galaxies before the mergers.
The assumption of bursty star formation does not apply to disc material accreted  after major mergers, which forms stars quiescently. Star formation in cusps and bars is always bursty.

\begin{figure}
\begin{center}
\includegraphics[width=0.99\hsize]{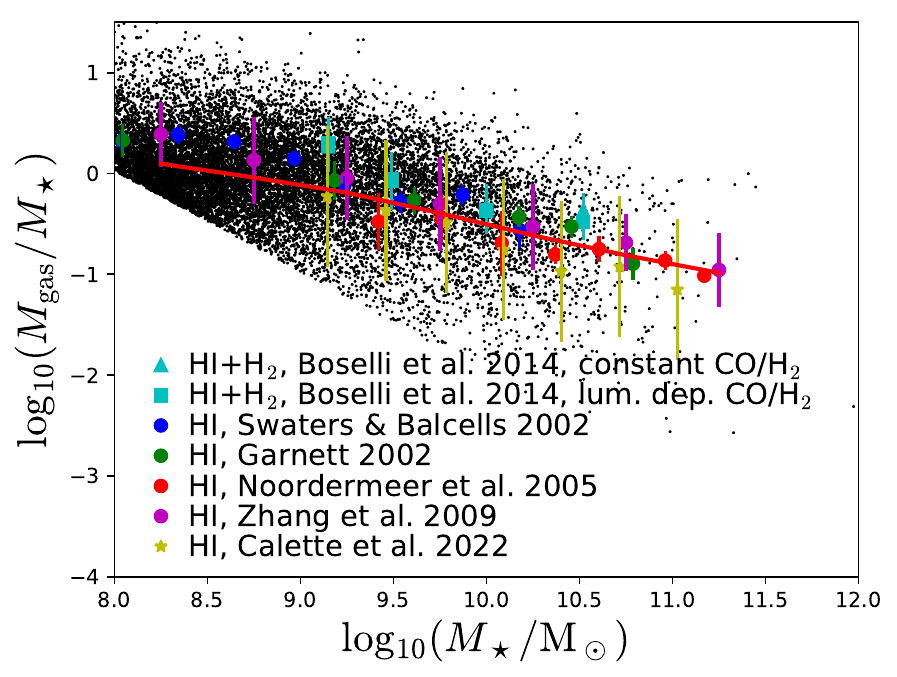} 
\end{center}
\caption{Gas-to-stellar mass ratio vs. stellar mass in GalICS~2.3 (black point clouds) and the observations (data points with error bars:  \citealp{boselli_etal14}; \citealp{swaters_balcells02}; \citealp{garnett02}; \citealp{noordermeer_etal05}; \citealp{zhang_etal09}; \citealp{calette_etal21}). The data points from \citet{boselli_etal14} are higher than those from the other authors because they include the molecular gas.
The model shown here includes tidal stripping, but the figure without tidal stripping is almost identical.
The black point cloud is for star-forming galaxies (sSFR$>10^{-11}{\rm\,yr}^{-1}$) with $M_{\rm gas}>10^8{\rm\,M}_\odot$.
The red line shows the mean relation in GalICS~2.3 with this selection criterion.
}
\label{GasFrac}
\end{figure}

In \citet{koutsouridou_cattaneo22}, the star formation timescales for quiescent and bursty star formation were  1\,Gyr  and 0.2\,Gyr, respectively.
That assumption had the advantages of being simple and robust to errors in disc sizes but is not the most physical description of star formation in discs.
The star formation timescale in the starburst mode is less important because it has no impact on the statistical properties of galaxies considered in this article as long as it is short compared with the age of the Universe.
Here we use a starburst star-formation timescale of 0.24\,Gyr (consistent with  \citealp{kennicutt_delosreyes21}) for both central and extended starbursts (i.e. starbursts in cusps and discs, respectively).

In this article, we revert to the original model of \citet{cattaneo_etal17}  for the quiescent mode and compute the star formation rate (SFR) by assuming that:
\begin{equation}
{\rm SFR}=\epsilon_{\rm sf}{M_{\rm gas}\over t_{\rm dyn}},
\label{sfr_disc}
\end{equation}
where $M_{\rm gas}$ is the mass of the gas in the disc,   $\epsilon_{\rm sf}$ is the efficiency of quiescent star formation, and
\begin{equation}
t_{\rm dyn}={2\pi r\over v_{\rm c}(r)}
\label{t_orb}
\end{equation}
is the orbital time at some characteristic radius $r$ (in our SAM, the exponential scale-length). The only difference with  GalICS~2.0 \citep{cattaneo_etal17} is that GalICS~2.3 does not contain any gas surface density threshold for star formation. That is based
on  \citet{delosreyes_kennicutt19}, who argue that, unlike the Schmidt law $\Sigma_{\rm SFR}\propto \Sigma_{\rm gas}^k$, the Silk-Elmegreen law $\Sigma_{\rm SFR}\propto \Sigma_{\rm gas}/t_{\rm dyn}$ has no clear turnover at low gas densities ($\Sigma_{\rm SFR}$ and $\Sigma_{\rm gas}$ are the SFR surface density and the gas surface density, respectively).

Eq.~(\ref{sfr_disc}) shows that  $\epsilon_{\rm sf}$ is degenerate with respect to $M_{\rm gas}$. That is the reason why our predictions for stellar masses and SFRs  are remarkably insensitive to small variations of $\epsilon_{\rm sf}$. If we increase $\epsilon_{\rm sf}$, we consume gas more rapidly, but the higher star formation efficiency compensates for the lower mass of the remaining gas, so that the final SFR is the same. The only way to break the degeneracy is to look at the gas content of galaxies. That is only possible in the local Universe because there are no {\sc Hi} data beyond $z\sim 0.1$.

In \citet{cattaneo_etal17}, we had calibrated  $\epsilon_{\rm sf}$ on the data of \citet{boselli_etal14} and found $\epsilon_{\rm sf}\sim 1/25$.
The main difficulty with that calibration was that the rotation speeds in \citet{boselli_etal14} were measured from {\sc Hi}  half line-widths rather than resolved rotation curves.
Hence, it is difficult to know at what radius they correspond. A simple solution is to assume a flat rotation curve.
We have also considered an alternative calibration based on \citet{delosreyes_kennicutt19} but here there is the additional complication that \citet{delosreyes_kennicutt19}
computed $t_{\rm dyn}$ at the outer radius of the star formation region, i.e., the radius $r$ that contains $\sim 95$ per cent of the {\sc H}$\alpha$ flux.
The star formation radius can be as small as half the exponential scale-length or as large as three exponential scale-lengths.
For an average star formation radius of 1.6 exponential scale-lengths, the data of \citet{delosreyes_kennicutt19} suggest $\epsilon_{\rm sf}\sim 1/15$. 
The two calibrations give an average timescale for quiescent star-formation at $z=0$ of $3.5\,$Gyr and $2.1\,$Gyr, respectively.
We have run GalICS~2.3 for both $\epsilon_{\rm sf}=1/25$ and $\epsilon_{\rm sf}=1/15$. The results are quite similar but slightly better for  $\epsilon_{\rm sf}=1/25$,
especially for the {\sc Hi} content of local galaxies (Fig.~\ref{GasFrac}). We therefore retain $\epsilon_{\rm sf}=1/25$ as our default model, also for consistency with our previous work.

We have verified that these changes in our model for quiescent star formation have no impact  on our previous results \citep{koutsouridou_cattaneo22}. The reason is the one discussed above. Moderate changes of the star formation timescale have almost no effect on stellar masses and SFRs. They are detectable only through their impact on the gas content of galaxies.

\subsection{Tidal stripping}

We model tidal stripping as in \citet{koutsouridou_cattaneo19} and  \citet{tollet_etal17}. The latter had followed a semi-empiral approach based on abundance matching rather than the semi-analytic one presented here. We find the tidal radius $r_{\rm t}$ of a satellite galaxy  by solving the equation:
\begin{equation}
{M_{\rm s}(r_{\rm t})\over r_{\rm t}^3}=\epsilon_{\rm ts}{M_{\rm h}(R)\over R^3}
\end{equation}
where $M_{\rm s}(r)$ is the satellite's mass within radius $r$ from the centre of the satellite, $R$ is the distance of the satellite from the centre of the host, $M_{\rm h}(R)$ is the mass of the host system within radius $R$ from its centre, and $\epsilon_{\rm ts}$ is the efficiency of tidal stripping. $M_{\rm s}(r_{\rm t})/r_{\rm t}^3$ is a decreasing function of $r_{\rm t}$. The higher $\epsilon_{\rm ts}$, the smaller $r_{\rm t}$.
The tidal radius is minimum when the satellite galaxy is at its orbital pericentre within the host system ($R=R_{\rm p}$).

We have recently discovered that \citet{tollet_etal17} had made a mistake of a factor of two in the calculation of $\epsilon_{\rm ts}$ while passing from their Eq.~(16)
to their Eq.~(17). 
The correct formula is:
\begin{equation}
\epsilon_{\rm ts}={\alpha^2\over -2\phi_{\rm s}(r_{\rm t})/v_{\rm c}^2(r_{\rm t})-1}\left[{V_{\rm c}(R_{\rm p})\over V_{\rm p} }\right]^2,
\label{epsilon_ts}
\end{equation}
where $\phi(r)$ is the gravitational potential of the satellite, $v_{\rm c}(r)$ is the circular velocity of the satellite, $V_{\rm c}(R)$ is the circular velocity of the host and $\alpha$ is a parameter of the model. 

Eq.~(\ref{epsilon_ts}) was derived assuming that the density distributions of the host and the satellite are power laws of $R$ and $r$, respectively, and that they have the same exponent $\alpha$.
We made that simplification so that we could tackle the problem analytically.
For the profile of \citet[NFW]{navarro_etal97}, the exponent varies from $\alpha=-3$ at large radii to $\alpha=-1$ in the inner region.
Numerical experiments where the host and the satellite follow NFW profiles are in reasonably good agreement with simple analytic estimates for $\alpha=-3$
\citep[Appendix~A]{tollet_etal17}, which, incidentally, corresponds to the classical Jacobi limit. 
Hence, we choose $\alpha=-3$ as our default value for models with tidal stripping.

\citet{tollet_etal17} thought they were computing tidal stripping for $\alpha=-3$ but they were effectively using a lower value (by a factor of  $\sqrt{2}$) because of the mistake in their Eq.~(17).
The model was in reasonably good agreement with the mass functions of central and satellite galaxies, but, on a closer look, the green curves in Fig.~13 of \citet{tollet_etal17} overestimated the stellar masses of  galaxies by at least $ 0.1\,$dex, which is logical, since their stripping was not strong enough. 
 
All figures in this article are for $\alpha=-3$. We have also run our model with $\alpha=0$ (no tidal stripping) for comparison. We found very little difference at $z>0.5$. At $z<0.5$, the model without tidal stripping displays an excess of  galaxies with $M_\star\sim 10^{12}{\rm\,M}_\odot$. They are brightest-cluster galaxies (BCGs) that grow to very large masses by cannibalising satellite galaxies. Tidal stripping 
{transfers some of the stellar mass of satellite galaxies to the intracluster light (ICL) before satellites merge with the BCGs  of their host holes. It thereby reduces the masses of BCGs by 0.1\,dex on average, in agreement with previous findings by 
\citet{tollet_etal17}. \citet{montes22}  finds the ICL of groups and clusters accounts for 50 to 70 seventy per cent of the total light of the ICL and the BCG combined. Assuming that stellar mass is proportional to luminosity, 
our finding is consistent with hers if 10 to 25 per cent of the observed ICL light comes from satellite galaxies that no longer exist because they have merged with the BCG (10 and 25 per cent correspond to an ICL fraction of 50 and 70 per cent, respectively).
One should note, however, that} the fundamental picture is the same with or without tidal stripping, {and that none of the results of this article is sensitive to our model for tidal stripping.}

\begin{figure*}
\begin{center}
\includegraphics[width=0.99\hsize]{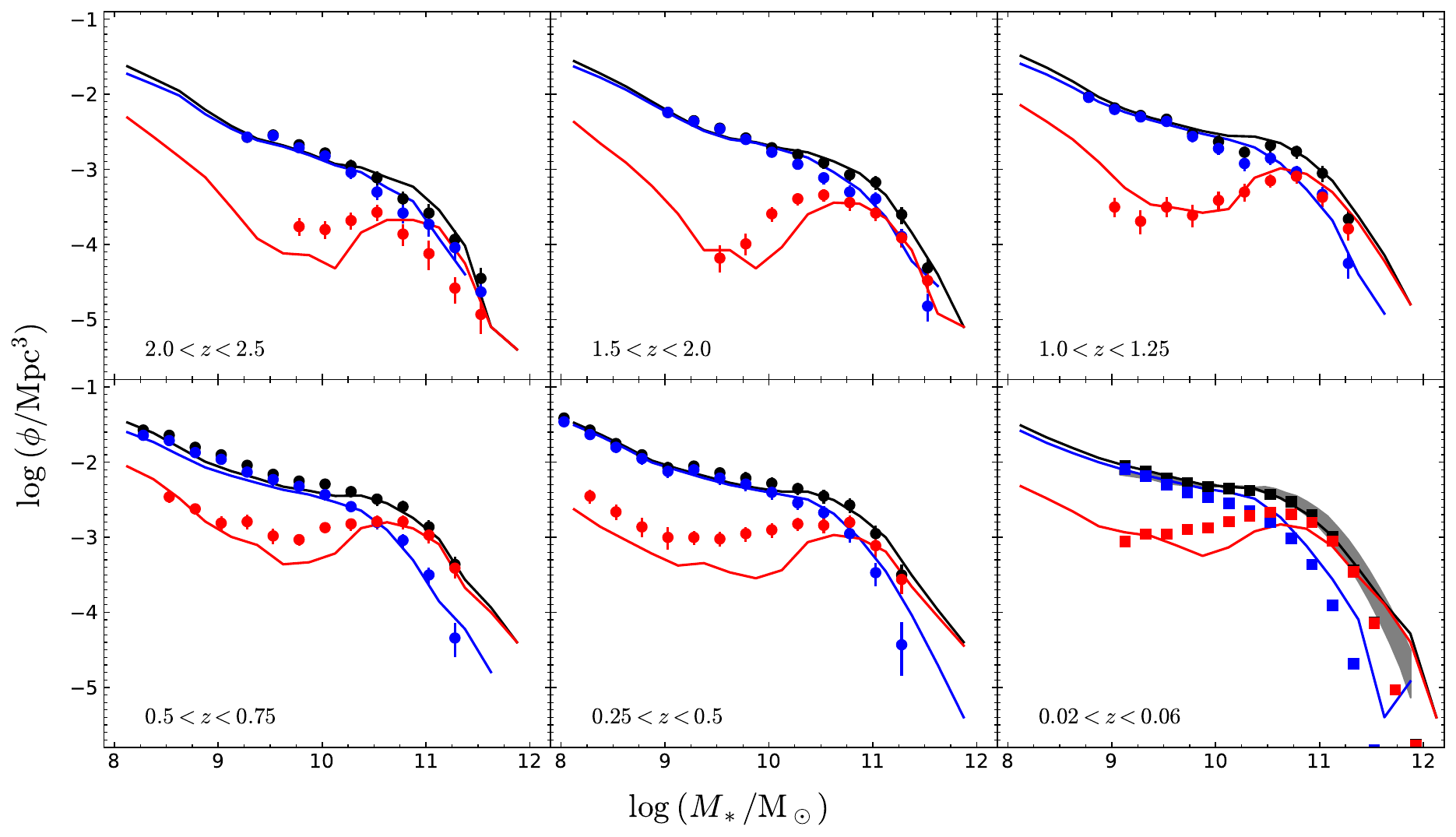} 
\end{center}
\caption{The mass functions of all galaxies ({\it black}), star-forming galaxies ({\it blue}) and passive galaxies ({\it red}). The curves show the predictions of GalICS~2.3.   The data points are the observations by  \citet{tomczak_etal14} at $0.2<z<3.0$ (circles) and \citet{weigel_etal16} at $0.02<z<0.06$ (squares). The gray shaded area in the panel at lowest $z$ shows
the possible range for the  mass function in the SDSS according to \citet{bernardi_etal17}.
The SAM has a formal mass resolution limit of $M_\star\sim 10^7{\rm\,M}_\odot$ (the stellar mass corresponding to the minimum halo mass that we can resolve based on the $M_\star$--$M_{\rm vir}$ relation in Fig.~\ref{HOD}).}
\label{MFs}
\end{figure*}

\begin{figure*}  
\begin{center}
\includegraphics[width=0.30\hsize]{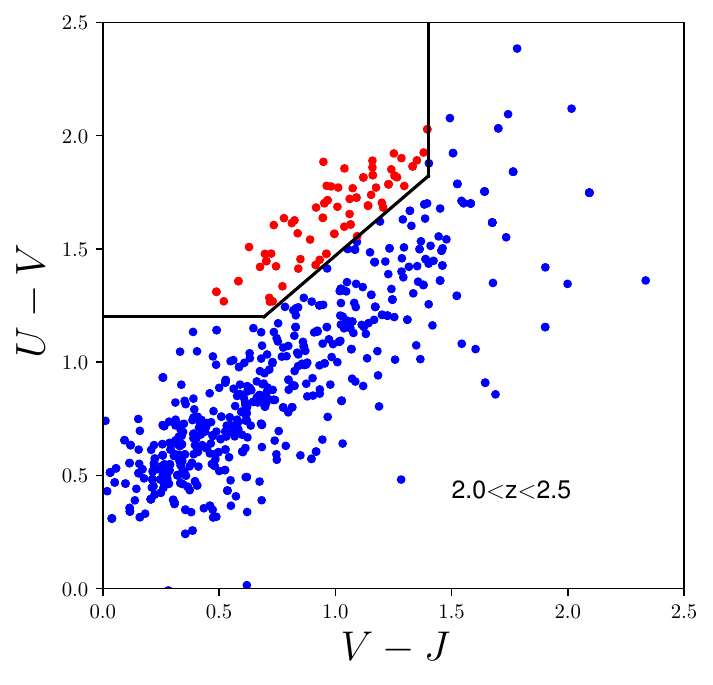} 
\includegraphics[width=0.31\hsize]{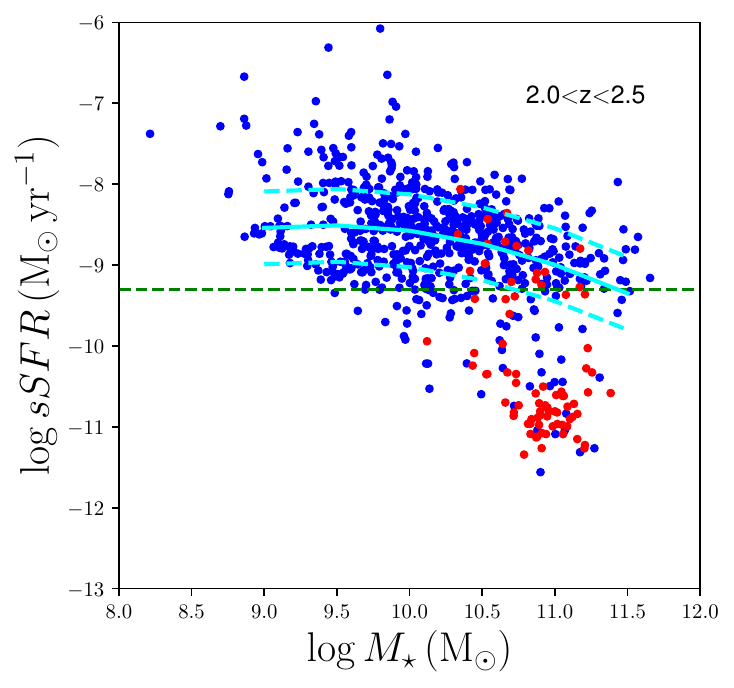} 
\includegraphics[width=0.31\hsize]{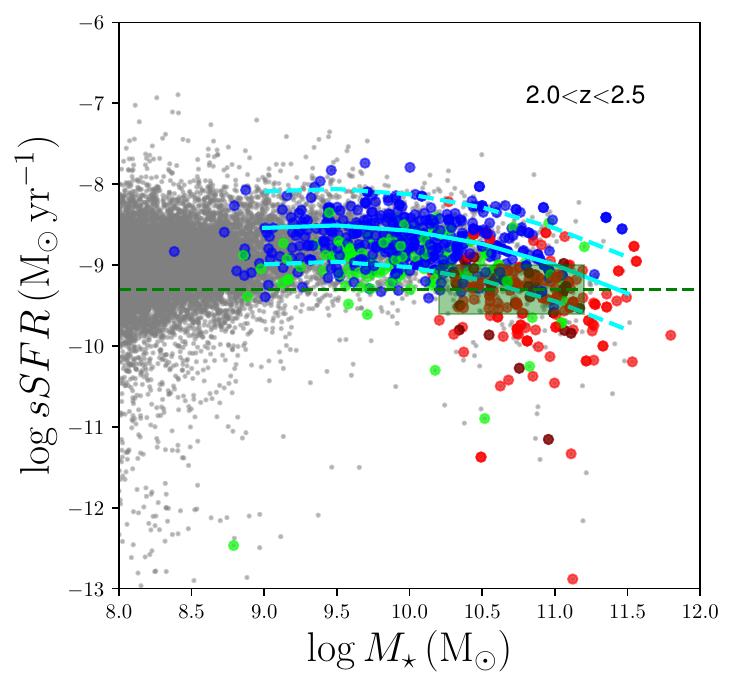} \\
\includegraphics[width=0.30\hsize]{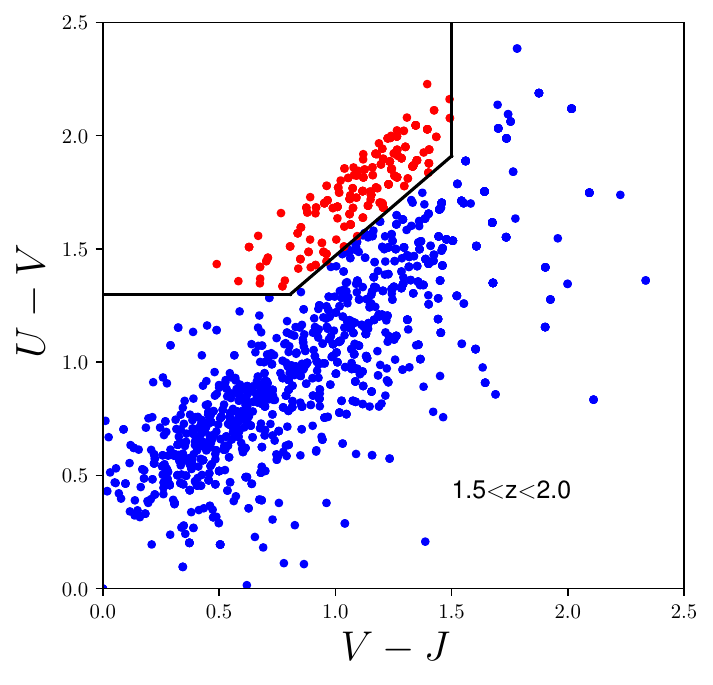} 
\includegraphics[width=0.31\hsize]{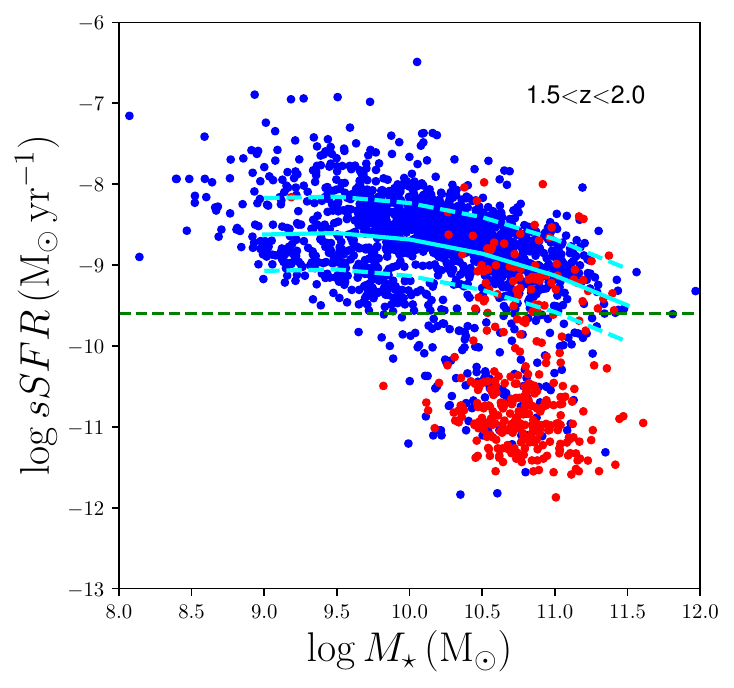}
\includegraphics[width=0.31\hsize]{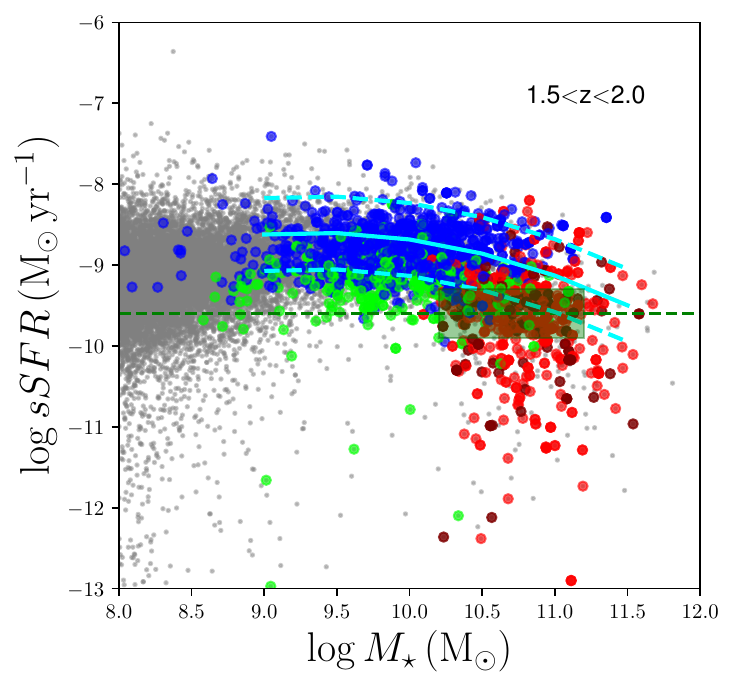} \\
\includegraphics[width=0.30\hsize]{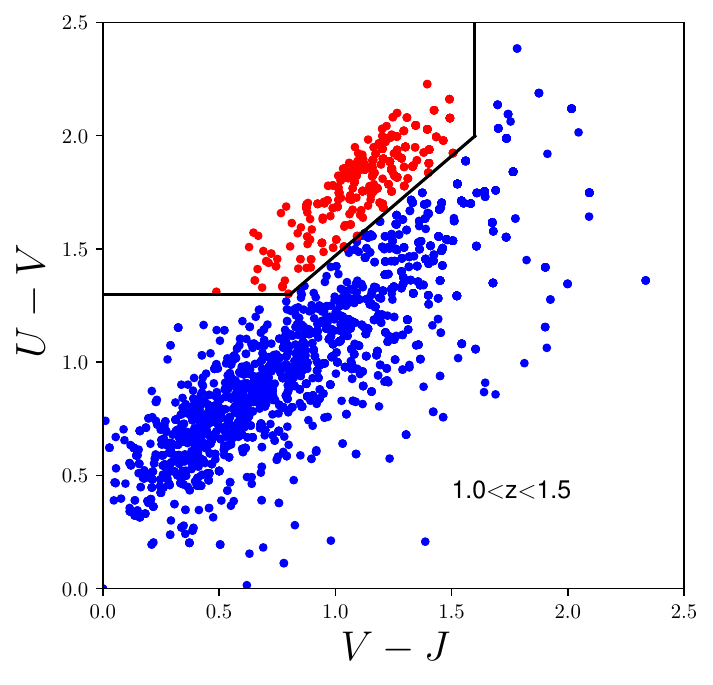} 
\includegraphics[width=0.31\hsize]{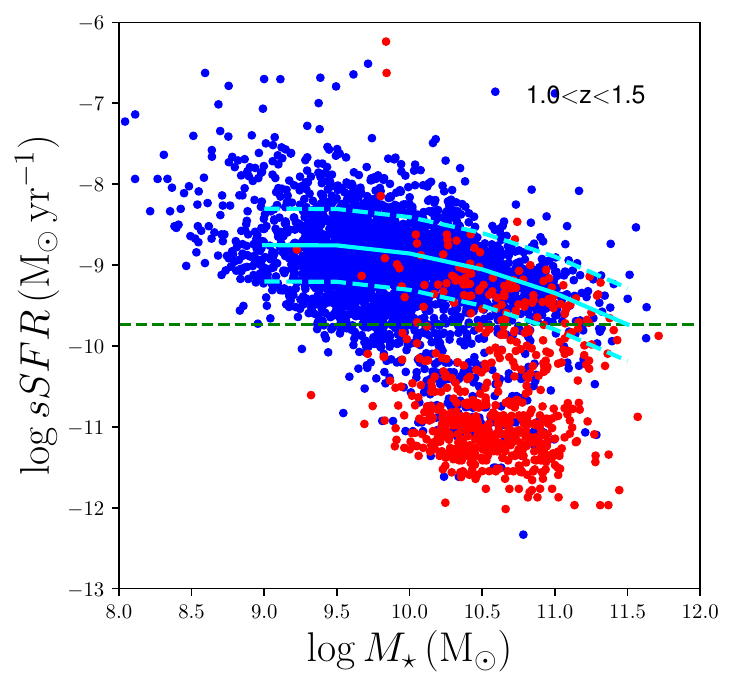}
\includegraphics[width=0.31\hsize]{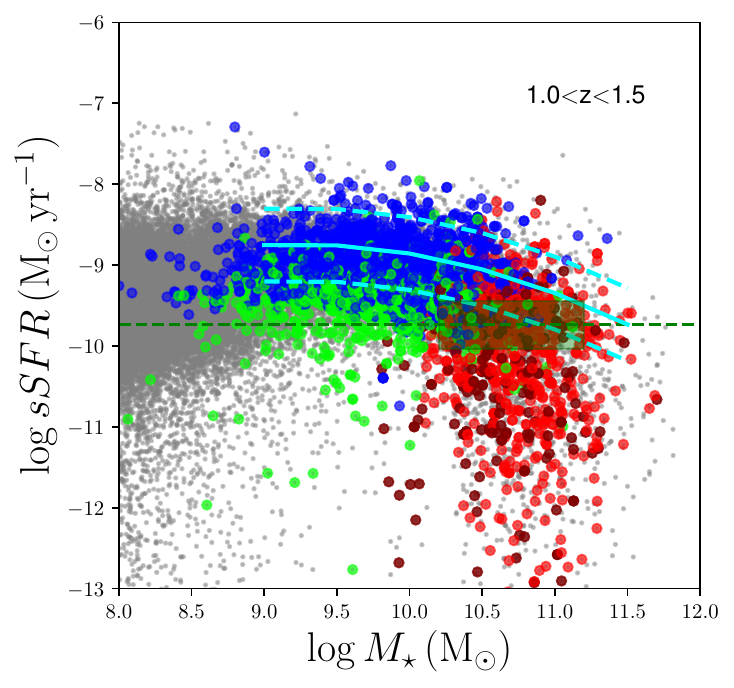} \\ 
\includegraphics[width=0.30\hsize]{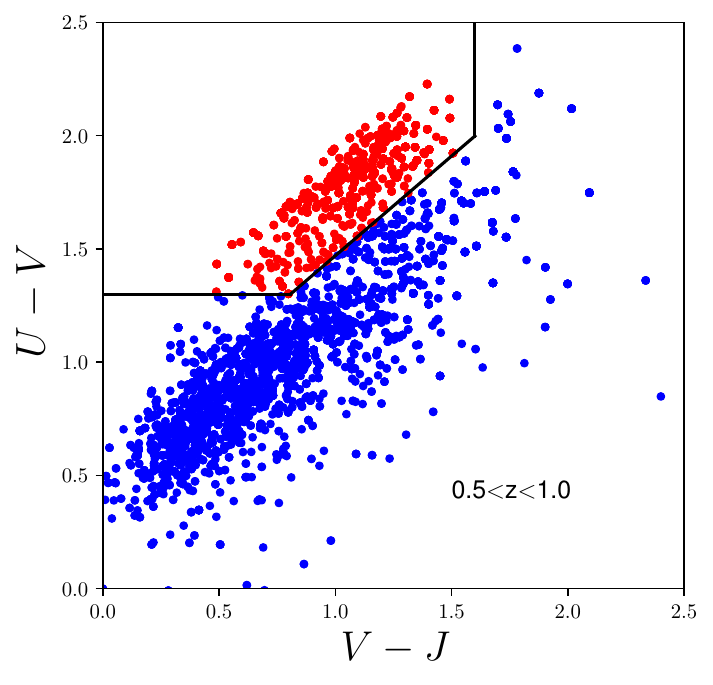}
\includegraphics[width=0.31\hsize]{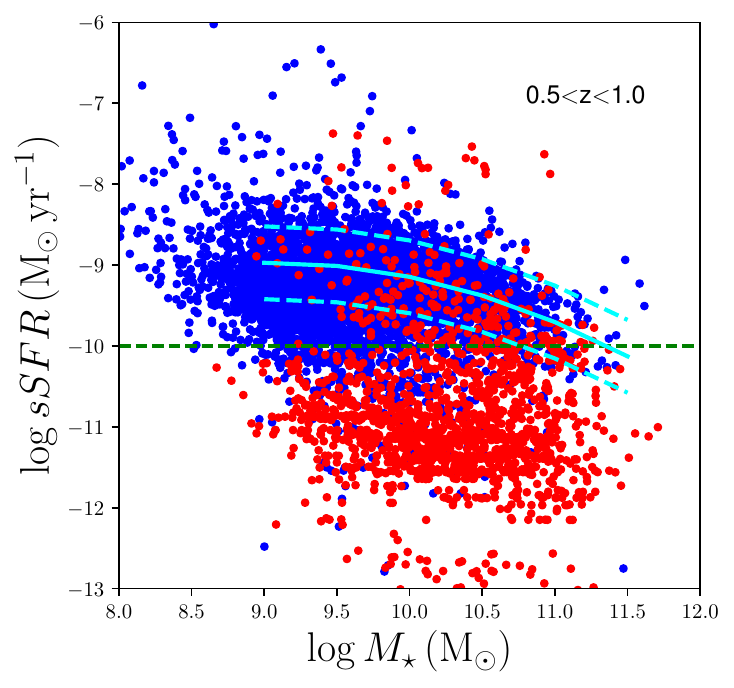}
\includegraphics[width=0.31\hsize]{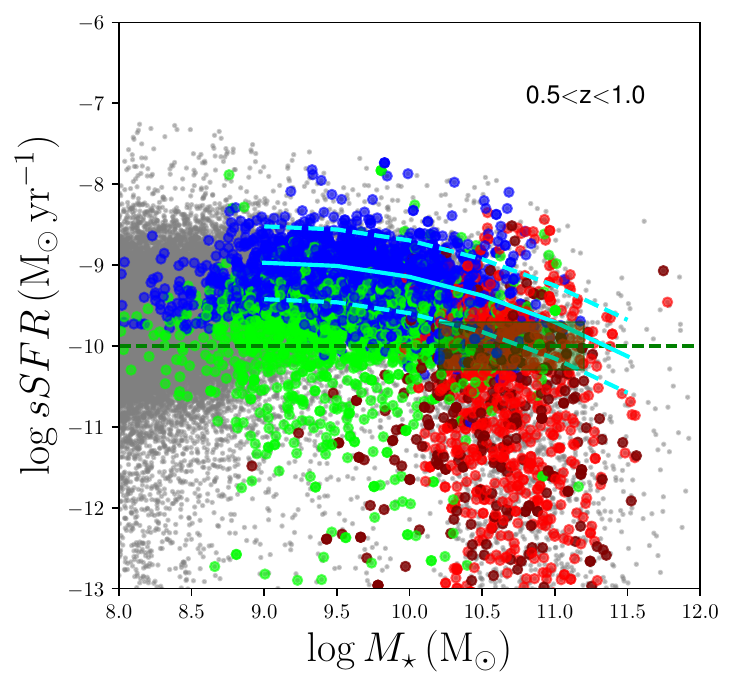} 
  
\end{center}
\caption{\citet{whitaker_etal11}'s colour-colour selection criterion in four bins of redshift (left) and the location of red and blue galaxies from CANDELS (classified according to this criterion) on a SSFR--stellar mass diagram (centre).
The right column shows the sSFR--$M_*$ in GalICS~2.3 (gray point cloud), which contains many more galaxies than CANDELS at low $M_\star$.
We have randomly selected a subsample of the gray points so that there is the same number of objects per interval of $M_\star$ in GalICS~2.3 as in CANDELS and we have coloured this subsample using four colours: 
blue for unquenched central galaxies, green for unquenched satellite galaxies, light red for quenched central galaxies, and dark red for quenched satellite galaxies.
The green dashes show the critical sSFR used to separate star-forming and passive galaxies.
The cyan curves show the  mean sSFR--$M_*$ of \citet{popesso_etal23} $\pm 0.45\,$dex. All lines are the same in the central and the right column to assist the comparison.
The green shaded rectangles on the right panels show that, for most galaxies, the transition from unquenched to quenched and from star-forming to passive occurs in the stellar mass range $10^{10.2}{\rm\,M}_\odot\lsim M_*\lsim 10^{11.2}{\rm\,M}_\odot$.
}
\label{CC}
\end{figure*}

\section{Mass functions}

In \citet{koutsouridou_cattaneo22}, we had already shown that our SAM reproduces the  mass function of galaxies at $0.5<z<2.5$ \citep{ilbert_etal13,muzzin_etal13,tomczak_etal14}
as well as the local data by \citet{yang_etal09}, \citet{baldry_etal12}, \citet{bernardi_etal13} and \citet{moustakas_etal13}. 
The novelty here is that we look at the mass functions of  star-forming and passive galaxies, separately. 
To compare our predictions with their data, we must consider how they separated the two populations in order to adopt a specific star formation rate (sSFR) threshold consistent with theirs.

\citet{tomczak_etal14}'s classification is based on \citealp{whitaker_etal11}'s 
redshift-dependent colour-colour (CC) criterion (Fig.~\ref{CC}, left panels).  The central column shows the {\it loci} of red and blue galaxies (according to \citealp{whitaker_etal11}'s CC criterion) from a CANDELS sample with SFR measurements on an sSFR--$M_\star$ diagram.
The sSFR distribution in CANDELS is bimodal. One can clearly distinguish a mainly blue high sSFR population and a mainly red low sSFR population.
The intermediate region with lower point density shifts to lower sSFRs at low $z$.
 
Our approach is to identify an sSFR boundary that separates star-forming   and passive galaxies in CANDELS  and  to apply that boundary to separate star-forming and passive galaxies in our SAM.
That approach makes sense if the main sequence of star-forming galaxies occupies the same sSFR range in GalICS~2.3 and CANDELS. If the sSFRs in GalICS~2.3 were systematically lower than those in CANDELS,
then a boundary based on CANDELS may run through our SAM's main sequence, so that many galaxies on our SAM's main sequence would be  misclassified as passive.

\begin{table}
\begin{center}
\caption{Critical sSFRs used to separate star-forming and passive galaxies.}
\begin{tabular}{cr}
\hline
\hline 
Redshift  & sSFR$_{\rm crit}\,({\rm yr}^{-1})$ \\
  \hline
$z<0.5$ & $10^{-11}$     \\            
$0.5<z<1.0$ & $10^{-10}$   \\
$1.0<z<1.5$ &$1.8\,10^{-10}$   \\  
$1.5<z<2.0$ & $2.5\,10^{-10}$ \\
$z>2.0$ &$5.0\,10^{-10}$\\
\hline
\hline
\end{tabular}
\end{center}
\label{model_parameters}
\end{table}

\begin{table*}
\begin{center}
\caption{Breakdown of the CANDELS sample by colour and sSFR  in five redshift intervals (numbers of galaxies).}
\begin{tabular}{crrrrr}
\hline
\hline 
Redshift   & Blue with sSFR$>{\rm sSFR_{\rm crit}}$    & Red with sSFR$>{\rm sSFR_{\rm crit}}$  & Blue with sSFR$<{\rm sSFR_{\rm crit}}$      & Red  with sSFR$<{\rm sSFR_{\rm crit}}$\\
  \hline          
$0.5<z<1.0$    &4384 & 252& 387 & 1327\\
$1.0<z<1.5$    &2852 & 110 &  325 & 537\\  
$1.5<z<2.0$  &1443 &  76  &    153 & 247\\
$2.0<z<2.5$  &597&  24  & 86 & 64\\
\hline
\hline
\end{tabular}
\end{center}
\label{model_parameters}
\end{table*}

\citet{popesso_etal23} combined the data from twenty-seven publications and fitted the main sequence with the cyan solid curves.
Expectedly, most of the star-forming in the CANDELS sample lie within $\pm 0.45\,$dex from \citet{popesso_etal23}'s main sequence (i.e. within the cyan dashed curves; Fig.~\ref{CC}).
If they did not, it would mean that our CANDELS sample is at odds with other observational studies.
The right panels shows how galaxies populate the sSFR--$M_\star$ diagram in GalICS~2.3. The colour of the symbols (discussed later) is not important for our current argument.
 The cyan lines are the same as in the middle panels. They have been replicated in the right panel to aid the comparison.
 Reassuringly, the bulk of the star-forming population in GalICS~2.3 lies within the cyan dashed curves, too.
 That means that we can indeed apply a sSFR criterion derived from CANDELS  to separate star-forming and passive galaxies in GalICS~2.3
 
 We have explored two ways of separating star-forming and quiescent galaxies on an sSFR--$M_\star$ diagram: horizontal cuts at redshift-dependent critical sSFRs  or cuts parallel the ridge line of the main sequence of star-forming galaxies (i.e. to the cyan curves). We found that  cuts  at the sSFRs in Table~1 (corresponding to the green horizontal dashed lines in Fig.~\ref{CC}) produced the best fit to the observational data (Fig.~\ref{MFs}).
 
 A word of caution is that  the comparison with the mass function of  \citet{tomczak_etal14}  is necessarily imperfect. The CANDELS data show the presence of red galaxies in the middle of the blue points on the main sequence of star-forming galaxies, so that no sSFR criterion can be entirely successful at reproducing the effects of a colour classification. Table~2  analyses this problem quantitatively. 
 For 86 per cent (at high $z$) to 90 per cent (at low $z$) of the objects in the CANDELS sample, a classification based on  \citealp{whitaker_etal11}'s CC criterion and one based on sSFR return the same outcome. In most cases, blue galaxies are star-forming and red galaxies are passive.
If the number of red galaxies incorrectly classified as passive 
and the number of blue classified as star-forming even though they have sSFR$<{\rm sSFR_{\rm crit}}$ were equal,
then these errors would cancel each other out and they would have no impact on the number densities of star-forming and passive galaxies.
Table~2 does show systematic differences, but the difference between the number of blue galaxies with sSFR$<{\rm sSFR_{\rm crit}}$  and the number of red galaxies with sSFR$>{\rm sSFR_{\rm crit}}$ (between the third and the second column)
constitutes  a small fraction of the total galaxy population at any given redshift: 8 per cent at $2<z<2.5$ and only 2 per cent at $0.5<z<1$.
Hence, the error that we commit by comparing predictions based on a sSFR criterion with observations  based on a CC criterion is unlikely to be significant.

The above conclusion is vindicated by comparing the mass functions of \citet{tomczak_etal14} at  $0.2<z<0.5$ with those by \citet{weigel_etal16} at $0.02<z<0.06$ (Fig.~\ref{MFs}).  \citet{weigel_etal16} did not use a CC criterion.
They had SFR measurements and they separated star-forming and passive galaxies at  sSFR$_{\rm crit}=10^{-11}\,({\rm yr}^{-1})$ as we do in our model at $z<0.5$. If an sSFR criterion and a CC criterion gave substantially different results,
there should have been substantial differences between red/blue squares at $0.02<z<0.06$ and the red/blue circles at $0.2<z<0.5$. We see none besides the normal evolution of the mass functions with redshift.

Fig.~\ref{MFs} shows that 
GalICS~2.3 is in excellent agreement with the observations at intermediate redshifts ($0.5<z<2.0$).
Its only weakness is a slight tendency to  underestimate the number densities of passive galaxies at intermediate masses (panels at $0.5<z<0.75$, $0.75<z<1$ and $1.5<z<2$).
 At $z>2$, GalICS~2.3 tends to overestimate the number densities of passive galaxies at high masses ($M_\star\gsim 10^{11}{\rm\,M}_\odot$) and to underestimate them at lower masses.
The discrepancy at $M_\star\gsim 10^{11}{\rm\,M}_\odot$  could be easily fixed by lowering sSFR$_{\rm crit}$ at $z>2$ but that would come at the price of an ever larger tension with the data at lower masses.
At low $z$, there is the opposite problem. At $M_\star\gsim 10^{11.5}{\rm\,M}_\odot$, GalICS~2.3 overpredicts the number densities of both star-forming and passive galaxies but especially of the former.
The problem manifests clearly at $0.02<z<0.06$, where the blue curve ends with a sudden upturn.

\citet{bernardi_etal17} have shown that the mass function of galaxies at $z\sim 0$  is much more uncertain than suggested by a purely statistical analysis because there are also systematic errors from the stellar population model and from the procedure used to fit the light profile. One cannot exclude that the mass function at $0.02<z<0.06$ may be as high at the upper boundary of the gray shaded area (Fig.~\ref{MFs}).
In that case, the discrepancy with the data points at $M_\star\gsim 10^{11.5}{\rm\,M}_\odot$ would derive from \citet{weigel_etal16}'s underestimating the stellar mass function at high $M_\star$.

Still, our SAM predicts the existence of star-forming galaxies with $M_\star\sim 10^{12}{\rm\,M}_\odot$, for which there are no observational counterparts.
We have focussed on the most massive galaxy in our computational volume, which happens to be star-forming, to investigate how that could be possible. 
We found that it has an anomalous growth history due to our model for orphan galaxies and ghost subhahoes
(Appendix~A).

 \begin{figure}
\begin{center} 
\includegraphics[width=0.99\hsize]{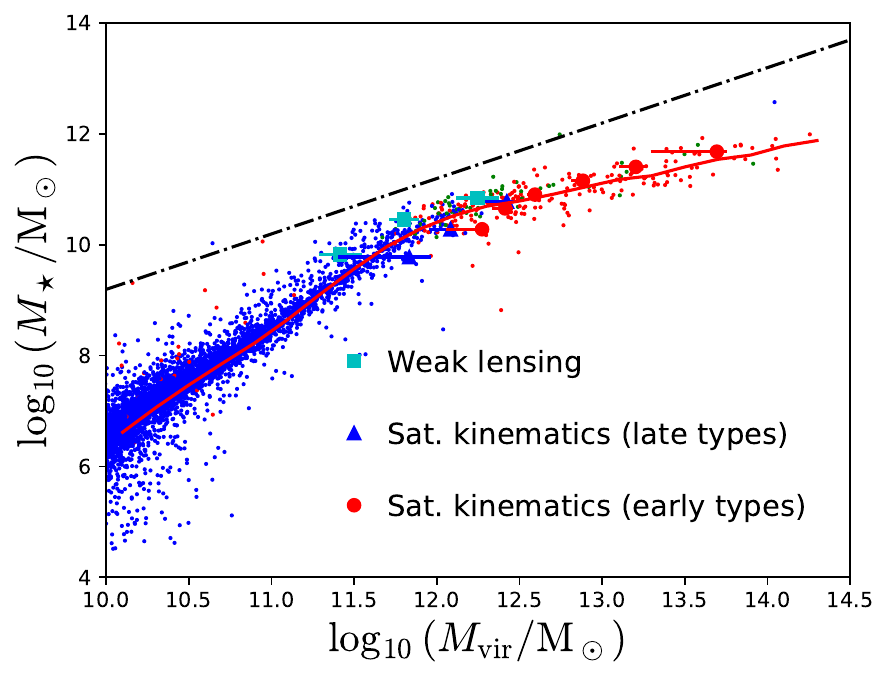} 
\end{center}
\caption{$M_\star$--$M_{\rm vir}$ relation in GalICS~2.3 at $z=0.05$: scatter plot (blue: unquenched galaxies; green: quenched star-forming galaxies; red: quenched passive galaxies) and median (red curve).
The symbols show the data points of \citet[cyan squares]{reyes_etal12} and \citet{wojtak_mamon13} for late-type (blue triangles) and early-type (red circles) galaxies.
To avoid overcrowding, we have shown only a subsample of the model galaxies. }
\label{HOD}
\end{figure}

Fig.~\ref{HOD} shows the local stellar--halo mass relation in GalICS~2.3.  The blue dots are ``unquenched" galaxies that have never satisfied Eq.~(\ref{quenching_crit}).
The red dots are quenched passive galaxies (galaxies with sSFR$<10^{-11}{\rm\,yr}^{-1}$). 
{The few green dots (corresponding to a number density of $0.675\times 10^{-3}{\rm\,Mpc}^{-3}$) are quenched galaxies with residual star formation.
GalICS~2.3 allows the  existence of such galaxies because in our SAM BH blows away all the gas in the central starburst but the gaseous disc is not blown away (Section~1). We have studied these objects in some detail.

 \begin{figure*}
\begin{center} 
\includegraphics[width=0.33\hsize]{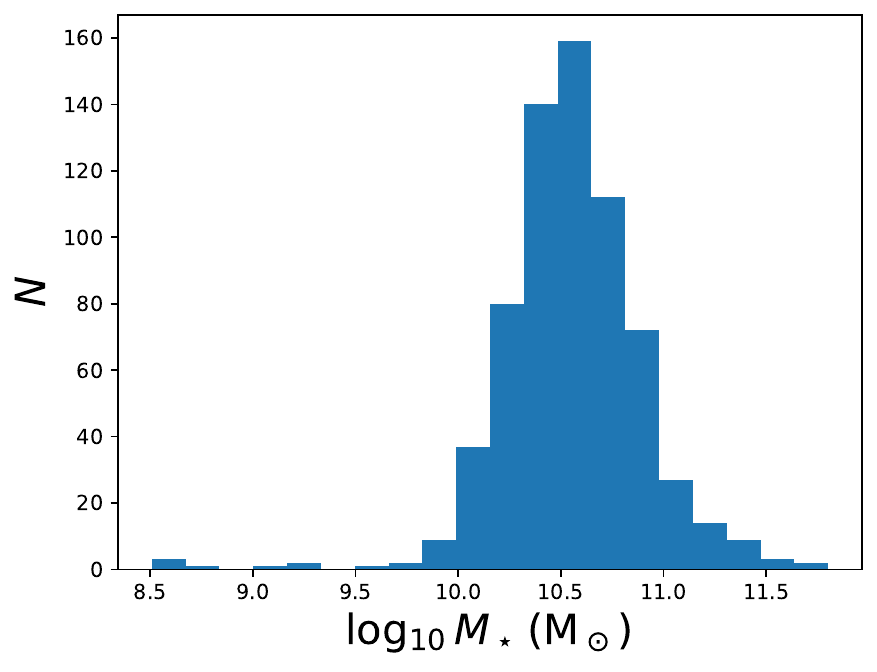} 
\includegraphics[width=0.33\hsize]{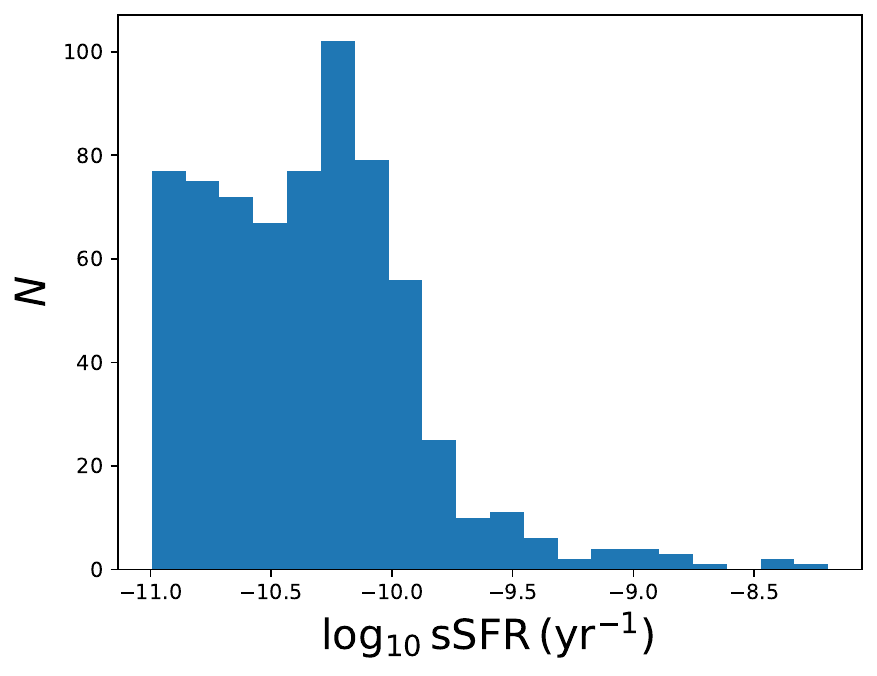} 
\includegraphics[width=0.33\hsize]{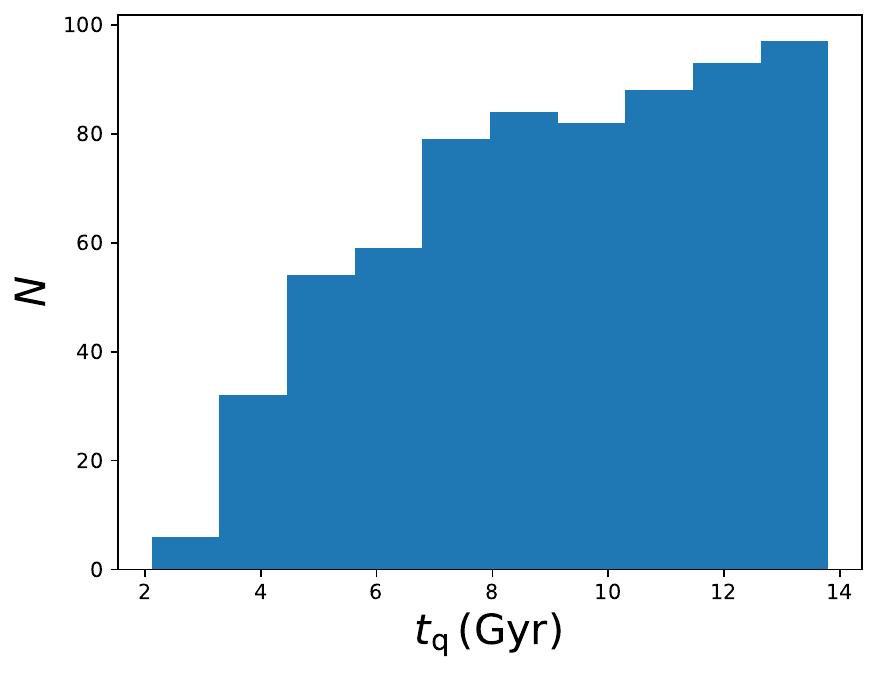} 
\end{center}
\caption{Distribution of stellar mass $M_\star$ (left), sSFR (centre) and quenching time $t_{\rm q}$ for quenched star-forming galaxies at $z=0.05$; $t_{\rm q}$ is the cosmic time (time since the Big Bang) at which quenching occurs.
$N$ is the number of galaxies in each bin (out of 675).}
\label{USF}
\end{figure*}

Quenched star-forming galaxies have a Gaussian mass distribution with $M_\star=10^{10.6\pm 0.3}\,{\rm M}_\odot$ (Fig.~\ref{USF}, left).
Their sSFR distribution has a peak at  $10^{-10.2}{\rm\,yr}^{-1}$ (Fig.~\ref{USF}, centre) corresponding to the ridge line of the main sequence of star-forming galaxies at $z=0$ for $M_\star=10^{10.6}\,{\rm M}_\odot$ but  is clearly skewed towards low sSFRs.
A little more than half of the quenched star-forming galaxies are still on the main sequence but many are entering or have entered the green valley. Very few have high sSFRs. 
They correspond to gas-rich major mergers, in which BH feedback has kicked in and all the gas is star bursting, whether in the cusp or the disc (Section~2.2).

The  distribution for the time of quenching $t_{\rm q}$ Fig.~\ref{USF}, right) shows that many star-forming galaxies experienced BH quenching several gigayears ago. 
To appreciate the significance of this finding, we compare
 $t_{\rm q}$  with the timescale on which a disc runs out of gas after it has ceased to accrete gas from the environment. In the local Universe,  a galaxy with sSFR$\sim 10^{-10.2}{\rm\,yr}^{-1}$ becomes  passive when its sSFR drops below $10^{-11}{\rm\,yr}^{-1}$ (Table~\ref{model_parameters}). For a gas fraction of 20 per cent (the gas fraction that corresponds to  $M_\star=10^{10.6}\,{\rm M}_\odot$; Fig.~\ref{GasFrac}), the sSFR drops by $0.8\,$dex in $1.7$ star-formation timescales, that is, $\sim 6\,$Gyr for a star-formation timescale of $3.5\,$Gyr (section~2.2).
That is an upper limit because, in our SAM, BHs grow because of mergers, and mergers always transfer some gas from the disc to the central cusp, where star formation is more efficient.
In fact, in major mergers, all the gas becomes starbursting with a depletion time of $0.24\,$Gyr (Section~2.2).
Hence the long  delay of $6\,$Gyr between the quasar phase and the actual shutdown of star formation is relevant only to galaxies that were quenched by a minor merger  and that have not experienced  any major merger 
thereafter.

Galaxies that were quenched more than 6$\,$Gyr ago (at $z>0.6$) cannot have kept making stars for such a long time without an external gas supply.
As we do not allow gas accretion onto quenched galaxies (Section~1), mergers are the only possible supply mechanism in our SAM.
Star-forming galaxies with $t_{\rm q}\lsim 10\,$Gyr are, in their majority, galaxies where mergers with gas-rich satellites have temporarily reactivated star formation.}

Fig.~\ref{HOD} shows that quenched passive galaxies dominate the $M_\star$--$M_{\rm vir}$ relation at high masses (the most massive galaxy clearly stands out as an outlier).
The red curve shows the median $M_\star$ as a function of $M_{\rm vir}$ in our SAM. GalICS~2.3 is in reasonably good agreement with the lensing data of \citet[cyan squares]{reyes_etal12} for local disc galaxies and with \citet{wojtak_mamon13}'s mass estimates from satellite kinematics for late-type (blue triangles) and late-type (red circles) galaxies.

GalICS~2.3 reproduces the upturn of the mass functions of passive galaxies at low $M_\star$, which is clearly visible in the data points at $0.25<z<0.75$.
This feature was already present in the luminosity function of early-type galaxies from the Two-degree Field Galaxy Redshift Survey \citep{folkes_etal99}, which was close to a Gaussian at high luminosities but rose again at low luminosities. More than twenty years ago,
\citet{cattaneo01} had remarked that merger remnants account for the Gaussian part at high luminosities but cannot explain the low-mass early-type population, for which one needs a different explanation.

\citet{ilbert_etal10}, \citet{tomczak_etal14} and \citet{huertas_etal15} have later confirmed the presence of two early-type populations: 
a massive bulge-dominated population with a Gaussian mass function centred on
$M_\star=10^{10.7\pm 0.5}\,{\rm M}_\odot$, which already existed at $z\sim 2$--3, and a lower-mass population that began to emerge much later.
Many of the galaxies in this second population are  red discs, which one may visually classify as lenticular because of the lack of spiral arms  \citep{huertas_etal15}.

In GalICS~2.3, these populations correspond to two pathways from the star-forming to the passive population: {BH quenching and environmental strangulation. To show the importance of these two mechanisms,
we have coloured the model galaxies in the right panels of Fig.~\ref{CC} according to both quenching status and their status as central or satellite galaxies: we have used blue for unquenched central galaxies, green for unquenched satellite galaxies, light red for quenched central, and dark red for quenched satellite galaxies.
Many galaxies are shown in gray, although they must fall in one of the four categories, because,  to have the same numbers of coloured symbols for the model and the observations, we have coloured only a subset of all model galaxies.}

In the first pathway (BH quenching), galaxies grow along the main sequence (cyan curves) until they reach the mass range where quenching occurs  ($10^{10.2}{\rm\,M}_\odot<M_\star<10^{11.2}{\rm\,M}_\odot$; \citealp{ilbert_etal10,huang_etal13,barro_etal17}). That is where the red symbols progressively replace the blue and green ones. Once quenched, galaxies drop out of the main sequence through the green rectangles, which correspond to the mass range above. This is the route through which the high-mass passive population was formed. This population is mainly composed of central galaxies (light red) although it also contains satellites galaxies (dark red).

The second pathway (environmental effects such as ram pressure and tides)  is more important at $M_\star<10^{11.2}{\rm\,M}_\odot$, 
where most of the galaxies with sSFR$<{\rm sSFR_{\rm crit}}$ are unquenched satellites (green circles).
{In GalICS~2.3, satellite galaxies  fall systematically below the main sequence of star-forming galaxies. At $z>1.5$, most of the green circles lie between the lower cyan dashed curve and  the green horizontal dashed line (below the lower boundary of the main sequence but still within the star-forming population). At $z<1.5$, however, a growing number of satellite galaxies start slipping below the green dashes, which are now running through the middle of the green symbols.
Interestingly, in CANDELS, too, there seem to be two populations of blue galaxies, one on the main sequence and one just below it (blue circles in the middle panels of Fig.~\ref{CC}; the presence of two populations is more obvious at $z>1.5$, where there are fewer data points). We have no information on the environment of these galaxies but we speculate that the blue CANDELS population with lower sSFRs may be the population of satellite galaxies, many of which are still making stars.

The finding that passive galaxies of low mass ceased to make stars because of environmental effects is consistent with chemical evidence that in low-mass galaxies star formation shut down slowly over several gigayears \citep{peng_etal15}. Also see \citealp{goubert_etal24} for evidence that quiescence in low-mass satellites correlates with the environment more than it does with BH mass.

Two considerations suggest  that GalICS~2.3 may not fully account for the importance of this second route. One is the deficit of passive galaxies  in our SAM at $M_\star\sim 10^{10}{\rm\,M}_\odot$ (Fig.~\ref{MFs}; most prominent at $0.2<z<0.5$), where environmental strangulation should be the main pathway to quiescence. The other is the excess of star-forming galaxies at high masses (most prominent at $0.02<z<0.06$). 
Most of these massive star-forming  galaxies  were quenched by BH feedback but have been replenished with gas by 
mergers with gas-rich satellite (Figs.~\ref{HOD} and \ref{USF}).
Perhaps the reason there are too many of them is that, in our SAM, ram-pressure and tidal stripping are not as effective as they should be.}

\begin{figure*}
\begin{center}
\includegraphics[width=0.45\hsize]{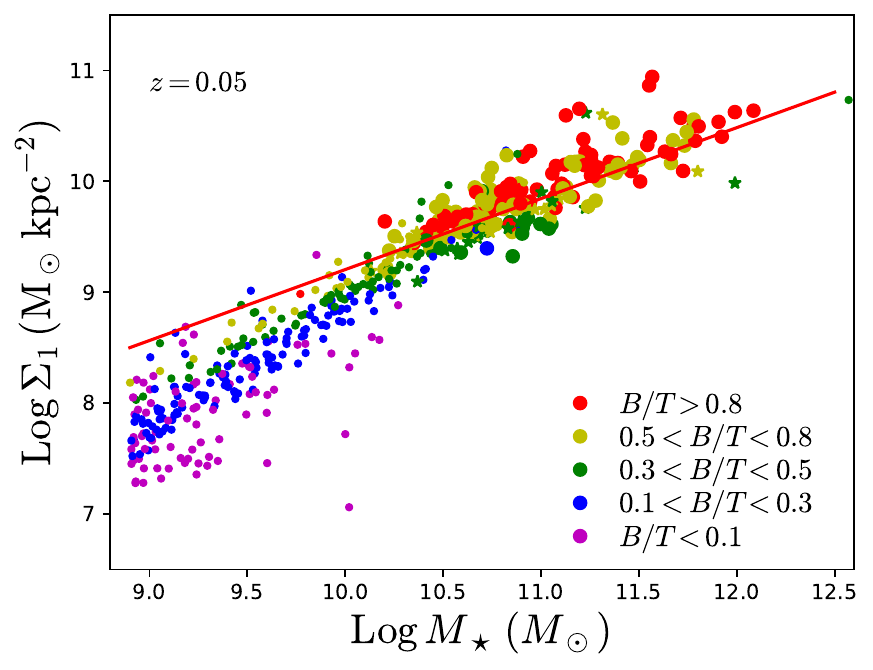} 
\includegraphics[width=0.45\hsize]{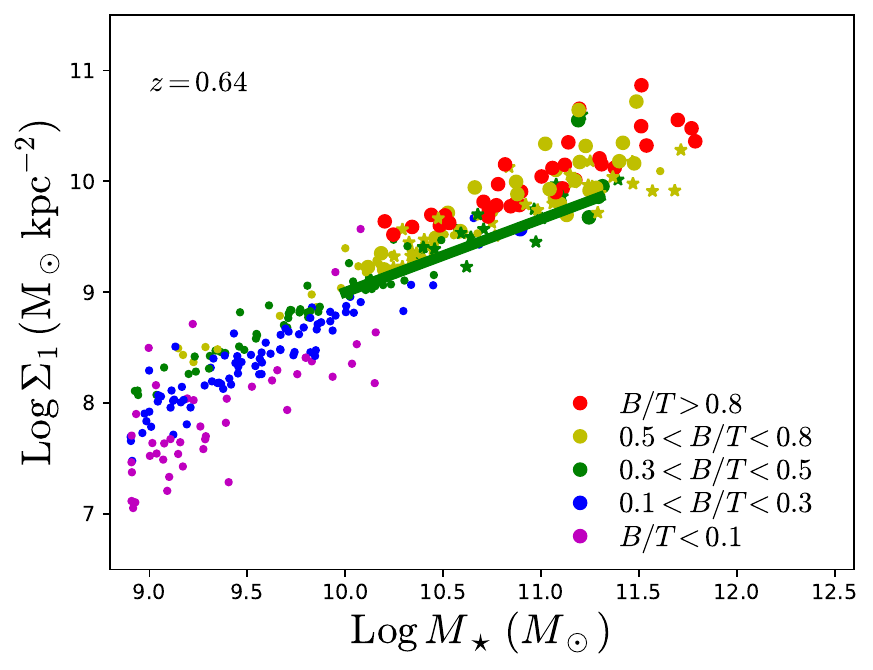}
\end{center}
\caption{$\Sigma_1$--$M_\star$ relation in GalICS~2.3 at $z\sim 0.05$ (left) and $z=0.64$ (right). The small circles are galaxies that have not been quenched (based on the quenching criterion in Eq.~\ref{quenching_crit}). 
The large circles are quenched passive galaxies.
The stars are galaxies that have been quenched but have not become passive yet because of residual star formation in the disc component.
Colour shows the bulge-to-total mass ratio $B/T$. 
The red line at $z=0.05$ corresponds to the ridge line of quiescent galaxies in ZENS  \citep{tacchella_etal17}. The green line at $z=0.64$ is \citet{chen_etal20}'s quenching boundary at that redshift.}
\label{tan}
\end{figure*}

\section{Surface densities}

In both CANDELS and GalICS~2.3, discs and bulges follow different mass--size relations.  For a same $M_\star$, 
bulges have smaller effective radii  than discs (Fig.~\ref{sizes}) and thus higher surface densities. Fig.~\ref{tan} shows that, in our SAM,
galaxies with higher bulge-to-total stellar mass ratios $B/T$ have statistically higher $\Sigma_1$. Unquenched galaxies (small circles) follow a different, steeper relation than quenched ones (large symbols).

\begin{figure*}
\begin{center}
\includegraphics[width=0.45\hsize]{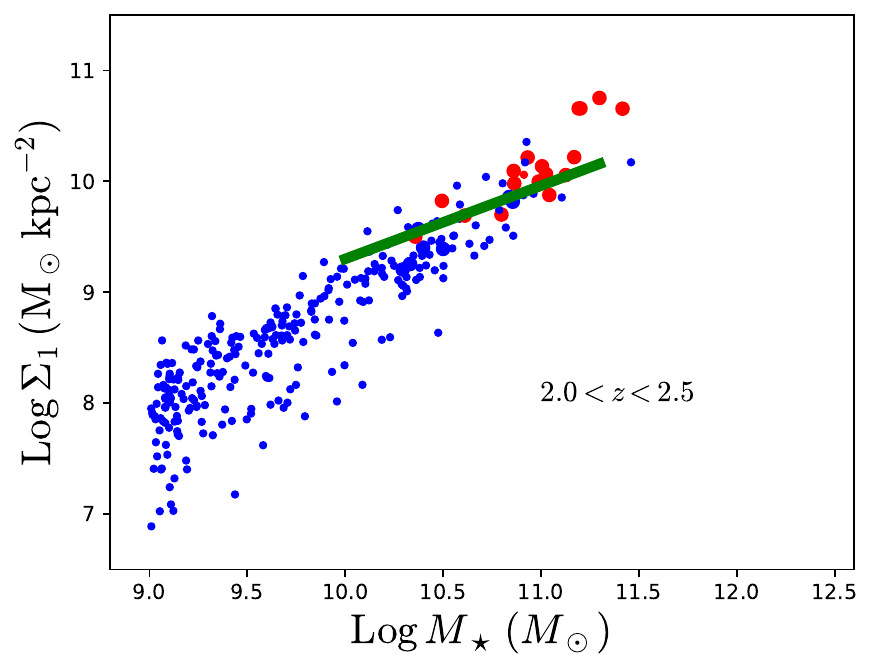} 
\includegraphics[width=0.45\hsize]{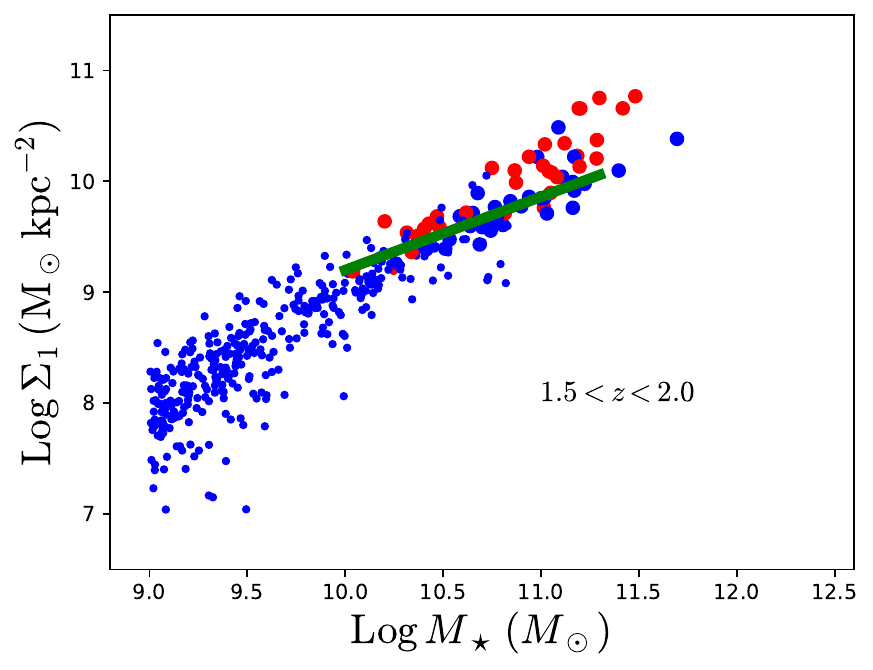}\\
\includegraphics[width=0.45\hsize]{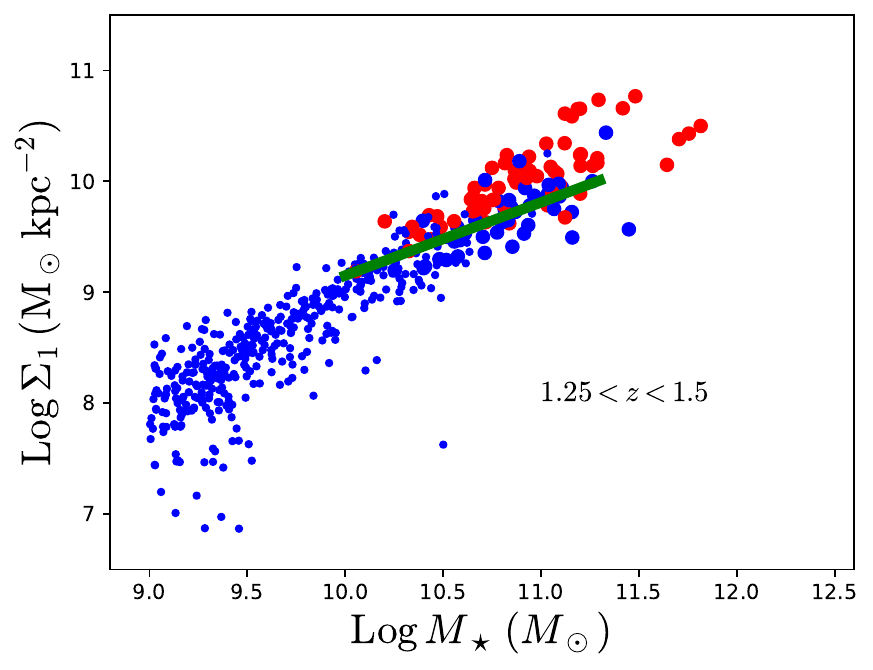} 
\includegraphics[width=0.45\hsize]{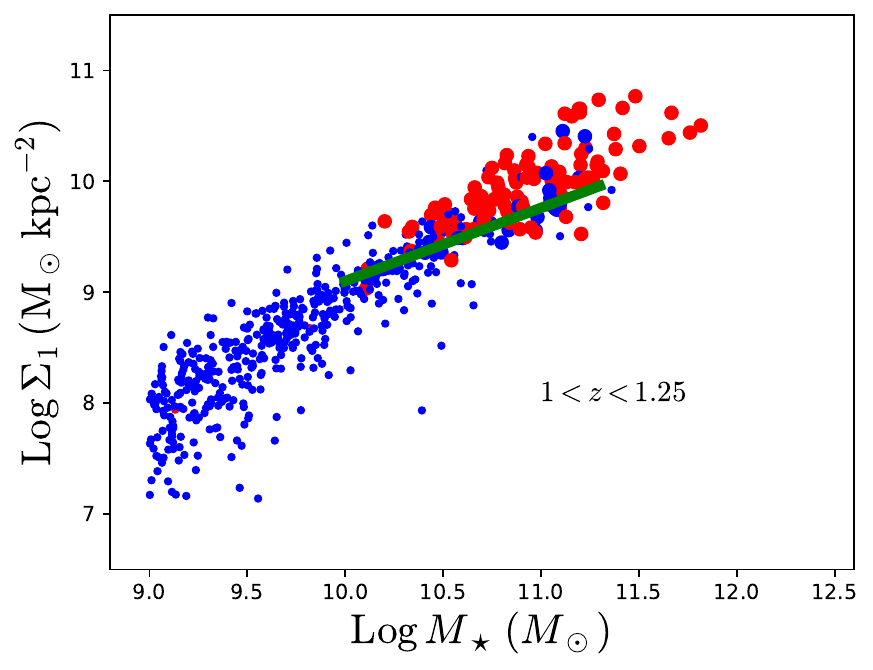}\\ 
\includegraphics[width=0.45\hsize]{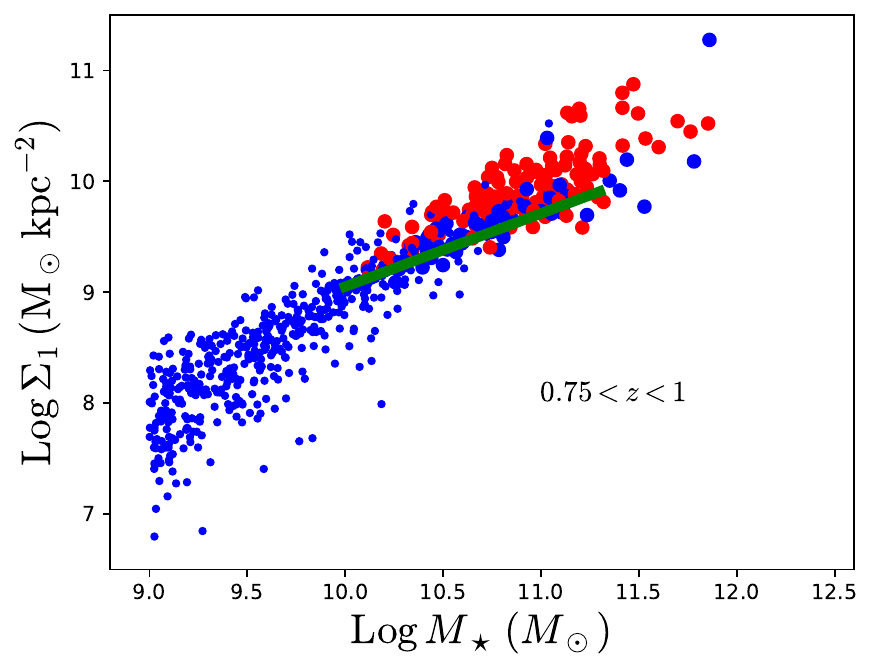}
\includegraphics[width=0.45\hsize]{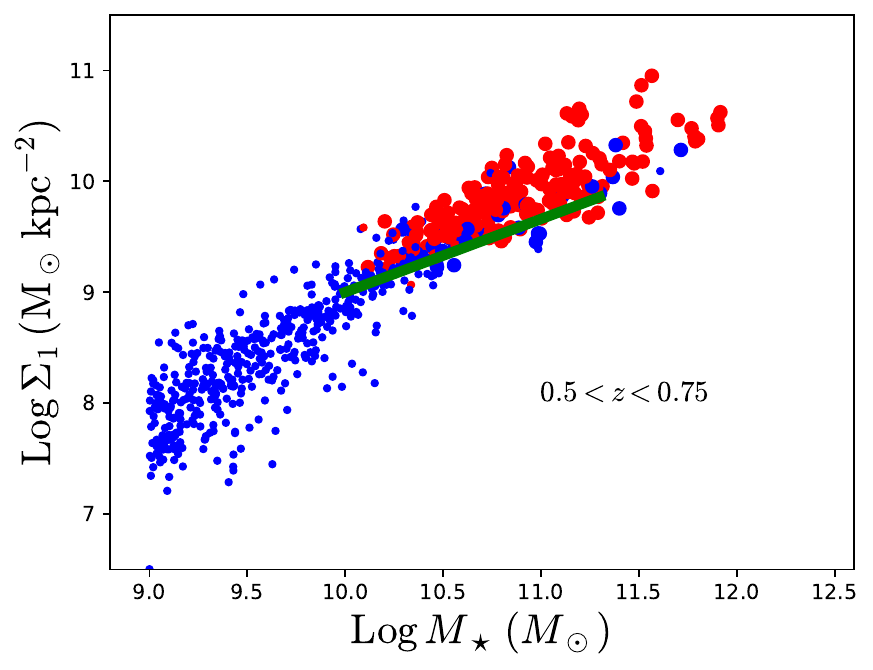}
\end{center}
\caption{$\Sigma_1$--$M_\star$ relation in GalICS~2.3 at $0.5<z<2.5$. Galaxies are shown with blue  or red circles according to whether they are star-forming or quiescent, respectively.
Large circles correspond to quenched galaxies, small circles to galaxies that have not been quenched. The green lines are the quenching boundaries of \citet{chen_etal20} in the mass range where they have been established observationally.
The population that ceased to make stars because of environmental effects (small red circles) does not appear on this figure because we have shown much fewer galaxies here than in Fig.~\ref{CC}.
If we had shown the same number of objects, the red sequence would be saturated by large circles.}
\label{Sigma1}
\end{figure*}

Fig.~\ref{tan} at $z=0.05$ is analogous to Fig.~2 of \citet{tacchella_etal17}, who plotted $\Sigma_1$ vs. $M_\star$ for galaxies in the Zurich ENvironmental Study (ZENS, $0.05<z<0.0585$)
and found a steeper relation for star-forming galaxies.
The red line is the  $\Sigma_1$--$M_\star$ relation for passive galaxies in ZENS and the SDSS \citep{fang_etal13}.
Fig.~\ref{tan} shows that  at $z=0.05$ quenched galaxies in GalICS~2.3  follow the SDSS/ZENS relation for passive galaxies.
The transition between the steeper relation for star-forming and the shallower relation for quenched galaxies occurs at the same mass scale as in \citet{tacchella_etal17}.

The difference, which we have already seen in Fig.~\ref{HOD}, is that in GalICS~2.3 not all quenched galaxies are passive. In Fig.~\ref{tan}, we have used stars to denote quenched galaxies with residual star formation and distinguish them from passive quenched galaxies (large circles). Once again, most quenched galaxies are passive, but Fig.~\ref{tan} does show seven star-forming galaxies on the $\Sigma_1$--$M_\star$ relation for passive galaxies at $M_\star>10^{11}\,{\rm M}_\odot$,
In  \citet{tacchella_etal17} there is only one despite the fact that our Fig.~\ref{HOD} and  Fig.~2 of \citet{tacchella_etal17} contain a similar amount of objects at $M_\star>10^{11}\,{\rm M}_\odot$.
With the exception of the most massive galaxy, which is anomalous and the only unquenched galaxy on the diagram at $M_\star>10^{11}\,{\rm M}_\odot$, the origin of the problem is that the discs of quenched galaxies keep forming stars for too long. Quenched ellipticals are all passive. The fraction of quenched galaxies with residual star formation is larger when there is a significant disc component.
In both ZENS and  GalICS~2.3, almost all star-forming galaxies lie below the red line.

\citet{tacchella_etal17}  coloured their data points by $B-I$. We have coloured our model galaxies by $B/T$. Their similarity is due to the colour--morphology relation.
Galaxies become redder as their morphologies evolves towards earlier types. Morphological evolution drives $\Sigma_1$ up. When $\Sigma_1$ crosses the quenching boundary, they leave the main sequence of star-forming galaxies and
migrate to the red sequence of passive galaxies.
In ZENS,  passive galaxies above the ridge line of the red sequence are systematically redder than those below it.
In their vast majorities, passive galaxies with $B-I\gsim 1.4$ are above the red line and
passive galaxies with $B-I\lsim 1.2$ are below it.
Fig.~\ref{tan} shows an analogous behaviour with respect to morphology. 
At a given $M_\star$, the average $B/T$ increases with $\Sigma_1$.
Most elliptical galaxies ($B/T>0.8$) are above the red line and they all correspond to quenched galaxies (large symbols).
Most of the galaxies in which star formation has been quenched are objects  with $B/T>0.5$.
Quenched galaxies with $0.1<B/T<0.3$ do exist (large blue symbols).
Their $\Sigma_1$ are lower than those of {quenched} galaxies with higher $B/T$, in the same way that not-so-red quiescent galaxies in ZENS have lower $\Sigma_1$ than redder objects.
Nearly all disc-dominated galaxies ($B/T<0.5$) are below the red line, independently of their quenching status.

In Fig.~\ref{Sigma1}, we extend our analysis of the $\Sigma_1$--$M_\star$ relation to the six redshift bins in Fig.~\ref{sizes}.
Galaxies are classified as star-forming or passive and shown in blue or red based on  sSFR (Table~1).
The green lines are the quenching boundaries of \citet{chen_etal20}. 
We have shown them in green as a visual reminder that they correspond to a sort of ``green valley" and that they
should not be confused with the ridge line of quiescent galaxies in Fig.~\ref{tan}, which we have shown in red because it corresponds 
to the red sequence.
 
The predictions of GalICS~2.3 are in good agreement with the CANDELS data.
The green lines mark the lower boundary of the passive population.
In \citet{chen_etal20}, almost all quiescent galaxies lie above the quenching boundaries. In Fig.~\ref{Sigma1}, almost all the red circles are above the green lines.
In \citet{chen_etal20}, most of the star-forming galaxies lie below the quenching boundaries, although there is a minority that lies above it.
The same applies to our model.

Quenched galaxies with residual star formation (the large blue circles above the green lines)  have lower $\Sigma_1$  than the red circles (passive galaxies).
That means that $\Sigma_1$ increases as the star formation fades away.
The conversion of gas into stars is not the main cause of the increase of $\Sigma_1$.
$\Sigma_1$ increases because of morphological evolution.
As we had already seen in \citealp[Fig.~10, panel~B2]{koutsouridou_cattaneo22},  $B/T$ increases with decreasing SFR not only as galaxies migrate from the star forming to the passive population
(i.e. to sSFR$<10^{-11}{\rm\,yr}^{-1}$ at $z\sim 0$) but also all the way down to  sSFR$=10^{-13}{\rm\,yr}^{-1}$.
This morphological evolution is due to further mergers between the one that triggered quenching and the final complete shutdown of star formation.
The quenched population is mainly composed of massive central galaxies (Fig.~\ref{CC}). Mergers play a major role in the mass assembly of these systems (e.g. \citealp{cattaneo_etal11}).
While it is certainly possible and even probable that the delay between quenching and quiescence may be too long in our SAM, we consider this feature to be robust. Galaxies evolve towards earlier morphological types (higher $B/T$ and $\Sigma_1$) as their stellar populations age after the episode that has triggered quenching.

\section{Summary and conclusion}

We have considered a simple model, in which major and minor mergers funnel
gas into the central regions of galaxies. Supermassive BHs feed on this  gas until the energy deposited into the ambient gas is enough to unbind it or to heat it to such high entropy that its accretion onto galaxies
is permanently shut down (``quenching").  In \citet{koutsouridou_cattaneo22}, we had focussed on the comparison with the local data. In this article, we have shown that our SAM:

1) is in good agreement with the evolution of the mass functions of star-forming and passive galaxies with redshift (especially at $0.5<z<2$, Fig.~\ref{MFs}), and 

2) reproduces \citet{chen_etal20}'s quenching boundary on the $\Sigma_1$--$M_\star$ diagram (Fig.~\ref{Sigma1}). 

The second finding is noteworthy because, although \citet{chen_etal20}'s empirical model had provided the initial motivation for our research, nowhere do the structural properties of galaxies enter our quenching criterion (Eq.~\ref{quenching_crit}).
Neither were any free parameters tuned to obtain the result in Fig.~\ref{Sigma1}, which is thus a genuine prediction of our model.

Our SAM's ability to reproduce the observations bolsters our confidence that we can use it to interpret them.
Our conclusion is that $\Sigma_1$ is a morphological indicator and that the quenching boundary arises because {\it the growth of supermassive BHs is linked to the formation of bulges}.
For a given stellar mass, galaxies with a larger $B/T$ have a larger BH mass.

In our interpretation, the quenching boundary on the $\Sigma_1$--$M_\star$ diagram is simply an aspect of the colour--morphology relation. 
The $\Sigma_1$--$M_\star$ relation is stratified in both colour/stellar age \citep{tacchella_etal17,luo_etal20} and $B/T$ (\citealp{luo_etal20} and Fig.~\ref{tan} of this article).
Spiral galaxies (low $\Sigma_1$) are mainly star-forming. Elliptical galaxies (high $\Sigma_1$) are almost all passive. The transition from late types to early types is both morphological and spectral.
The $\Sigma_1$ quenching boundary corresponds to the critical $B/T$ that separates the star-forming spiral population from the passive S0/elliptical population.
In our SAM, the critical $B/T$ is in the range $0.3<B/T<0.5$ at $z\lsim 1.3$ (Fig.~\ref{tan}, right panel) and slightly higher at earlier epochs, consistently with \citet{chen_etal20}'s finding
that the quenching threshold increases with $z$.
 
Fig.~\ref{tan} of this article and Fig.~10 of \citet{koutsouridou_cattaneo22} show  that  $B/T$ keeps increasing after the event that triggered quenching because of further mergers while the SFR fades away. This finding is 
important to explain why the $\Sigma_1$--$M_\star$ relation display colour stratification even within the passive population and is in agreement with \citet{cattaneo_etal11}, who estimated that galaxies with $M_\star>10^{11}\,{\rm M}_\odot$ accreted most of their mass through dissipationless mergers.

BH quenching explains the high-mass passive population. The mass function of this population has a maximum at $M_\star\sim 10^{10.7}\,{\rm M}_\odot$ (Fig.~\ref{MFs}).
The mass function of passive galaxies rises again towards low masses at $M_\star\lsim 10^{9}{\rm\,M}_\odot$. 
This low-mass passive population is almost entirely composed of satellite galaxies and is the product of environmental effects such strangulation, ram-pressure stripping and tidal stripping.
This finding is consistent with \citet{huertas_etal15}'s morphological analysis (the lower-mass passive population is dominated by red discs/lenticulars  rather than elliptical galaxies) 
and \citet{peng_etal15}'s chemical analysis (star formation has depleted the gas reservoir over a few gigayears
after accretion from the environment has stopped).

\section*{Data availability statement}

The results of these articles do not depend on any non-public data.


\bibliographystyle{mn2e}

\bibliography{ref_av}

\appendix

\section{Mass resolution of the merger trees}

The N-body simulation used to construct the merger trees employed $1024^3$ particles for a volume of $(100{\rm\,Mpc})^3$. For our cosmology that corresponds to an N-body particle mass of $3.7\times 10^7{\rm\,M}_\odot$.
The halo finder ({\sc HaloMaker}; \citealp{tweed_etal09}) detects haloes with more than a hundred particles. Previous comparison of the halo mass functions from N-body simulations with the analytical fit by \citet{sheth_etal01} have shown that are halo catalogues are complete above a mass limit of a few hundred particles (Fig.~1 of \citealp{cattaneo_etal17}). In our case, that corresponds to a {halo} mass of $M_{\rm vir}\sim 10^{10}{\rm\,M}_\odot$ and a stellar mass
of $M_\star\sim 10^7{\rm\,M}_\odot$ based on stellar--halo mass relation in Fig.~\ref{HOD} of this article.

Subhaloes are  much more difficult to resolve, especially when they are inside massive systems (i.e. groups and clusters of galaxies). 
A subhalo that the halo finder can no longer resolve is deemed to have merged with its host system.
That leads to the numerical overmerging of satellite galaxies with the central dominant galaxies of their host haloes.

To obviate this problem, we complete the {\sc HaloMaker} catalogues with ghost subhaloes, i.e. subhaloes that the halo finder no longer detects but that should still exist based on the model of \citet{tollet_etal17}, where we showed
that a halo catalogue constructed from simulation with $512^3$ particles and completed with ghosts is a good as one constructed from a simulation with $1024^3$ particles without ghosts in terms of its ability to reproduce the conditional mass functions of satellite galaxies in groups and clusters.

\begin{figure*}
\begin{center}
\includegraphics[width=0.99\hsize]{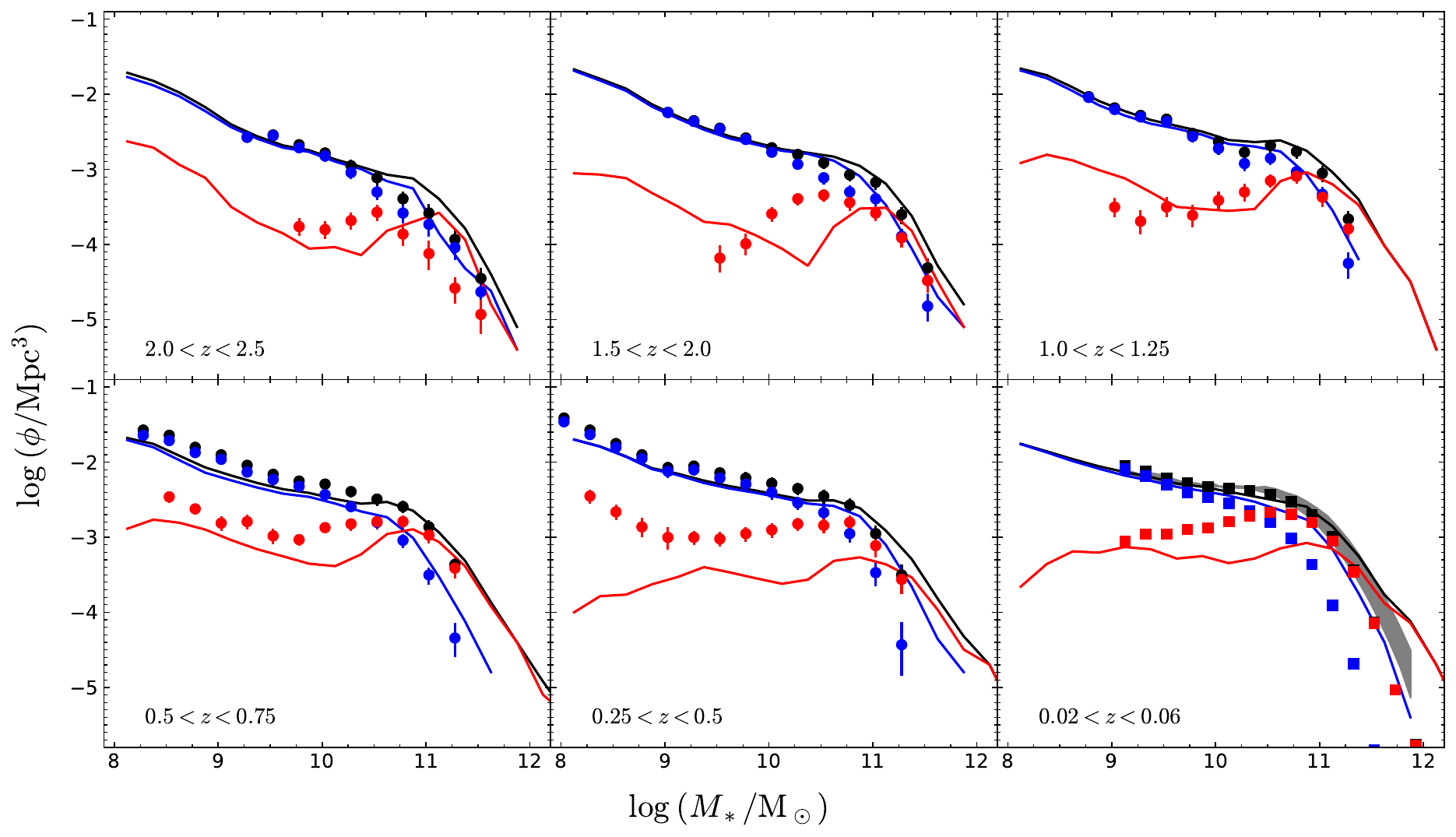} 
\end{center}
\caption{As Fig.~\ref{MFs} but for merger trees: 1)  from a simulation with $512^3$ instead of $1024^3$ particles and 2) without ghost subhaloes.
The model shown here has a stellar mass resolution of $M_\star\sim 10^8{\rm\,M}_\odot$ for central galaxies.
For satellite galaxies, the mass resolution is worse by at least an order of magnitude.}
\label{MFs_lowres}
\end{figure*}

To test how numerical resolution affects our results, we have rerun GalICS~2.3 on the merger trees of \citet{cattaneo_etal17} and used the results to recompute the mass functions of star-forming and passive galaxies (Fig.~\ref{MFs_lowres}).
 \citet{cattaneo_etal17} used a simulation with the same cosmology, the same volume and the same initial condition as ours but withy only employed $512^3$  particles and no ghost subhaloes.
 In Fig.~\ref{MFs_lowres}, the mass resolution is eight times lower than in Fig.~\ref{MFs} for haloes and far worse for subhaloes. Unsurprisingly, there are differences, which we are about to discuss, but they are not large.

At low masses,  the black curves are slightly lower in Fig.~\ref{MFs_lowres} than in Fig.~\ref{MFs}. For the red curves, the difference is much more pronounced.
 Expectedly, at lower resolution there are fewer dwarf galaxies and especially fewer passive dwarves, most of which are satellites and therefore poorly resolved. 

At high masses, the low-resolution simulation gives a much higher fraction of star forming galaxies. In Fig.~\ref{MFs_lowres}, the red and blue curves for $0.02<z<0.06$ cross at $M_\star\sim 10^{11.2}{\rm\,M}_\odot$.
In the observations, the fraction of star-forming galaxies at $M_\star\sim 10^{11.2}{\rm\,M}_\odot$ is $\sim 10$ rather than $\sim 50$ per cent.
At high resolution (Fig.~\ref{MFs}), the fraction of star-forming galaxies is still on the high side (around 20 per cent) but the discrepancy with the observations is much smaller.
The trend is reassuring and let us presume that the agreement with the observations would further improve if we could use merger trees from an N-body simulation with even higher resolution.

We have run several tests to investigate why higher resolution increases the fraction of passive galaxies.
At low resolution, galaxies of the same mass have lower $B/T$ ratios but higher {\it ex-situ} fractions (stellar mass fractions acquired through mergers).
That can happen only if the additional {\it ex-situ} stars come from minor mergers, which are less effective at increasing $B/T$ and growing supermassive BHs but can supply gas to the discs of massive galaxies.

Ram pressure and tidal stripping remove gas from satellite galaxies and thus limit the importance of this route to replenish quenched massive central galaxies with gas.
However, both ram pressure and tidal stripping  are more effective when satellite galaxies are closer to their orbital pericentres.
At low resolution many subhaloes merge with their hosts long before reaching their pericentres.
We have checked that, for a same set of DM merger trees, $B/T$ ratios decrease and {\it ex-situ} fractions increase when these processes are turned off.

There are two reasons why minor mergers become more important at low resolution. Both are connected to the fact that low resolution leads to overmerging (see above) and that degrading the resolution has a greater impact on galaxies with lower masses.
First, degrading the resolution increases mergers rates at high redshift, where halo masses are smaller, and that explains why the number densities of massive galaxies  at $2<z<2.5$ are higher at low resolution (compare Figs.~~\ref{MFs} and~\ref{MFs_lowres}). Once a galaxy has grown too much, it becomes hard for it to find companions of comparable mass later on. Hence minor mergers become more probable. 
Second, at all redshifts, the merging rates of the galaxies with lowest masses receive the greatest boost and that contributes to make minor mergers comparatively more important.

The most massive galaxy in Fig.~\ref{HOD} is unquenched because it was able to reach a very large mass without any major merger.
We have conducted several tests to investigate how that could possibly happen. We have discovered that this system has a pathological growth due to the algorithm that creates ghost subhaloes. 
The goal is to prevent premature merging but for this object some mergers are excessively delayed.

To investigate the extent to which pathologies of the merger trees may affect our results, we have rerun GalICS~2.3 on DM merger trees constructed using a different algorithm \citep[{\sc consistent trees}]{behroozi_etal13b}, a different halo finder \citep[{\sc rockstar}]{behroozi_etal13a}
and a different N-body simulation \citep[SMDPL]{klypin_etal16}.
Clearly, GalICS~2.3 perform better with its native merger trees, on which it was calibrated, and other merger trees have issues  of their own (see \citealp{diemer_etal23}).
However, the results were not substantially different. That bolsters our confidence  that the resolution of the N-body simulation and the algorithms used to extract merger trees from the simulation should not be a major cause of concern.

\end{document}